\documentclass[10pt, twocolumn, amsmath,amssymb,a4paper, pra]{revtex4}

\usepackage{theorem}
\newcommand{\ket}[1]{| #1 \rangle}
\newcommand{\bra}[1]{\langle #1 |}
\newcommand{\one}{\mathrm{I} \! \! 1}

\newcommand{\B}{\mathfrak{B}}
\newcommand{\Hi}{\mathcal{H}}

\newcommand{\Tr}{\mathrm{tr}}
\theorembodyfont{\rmfamily}
\newtheorem{Theorem}{\textit{Theorem} }

\newtheorem{Definition}{\textit{Definition} }
\newtheorem{Corollary}{\textit{Corollary}}
\newtheorem{Lemma}{\textit{Lemma}}
\newtheorem{Proof}{\textit{Proof}}

\begin{document}

\title{Entanglement of multiparty stabilizer, symmetric, and antisymmetric states}

\author{Masahito Hayashi$^{1,2}$,  Damian Markham$^{3,4}$, Mio Murao$^{3,5,6}$, Masaki Owari,$^{3,6}$ and Shashank
Virmani$^{7,8}$}

\address{$^1$Graduate School of Information Sciences,
Tohoku University,
Aoba-ku, Sendai, 980-8579, Japan \\
$^2$ERATO-SORST Quantum Computation and Information Project
Japan Science and Technology Agency, \\
201 Daini Hongo White Building.
5-28-3, Hongo, Bunkyo-ku, Tokyo 113-0033, Japan
\\
$^3$Department of Physics, Graduate School of Science,\\
 The University of Tokyo, Tokyo 113-0033, Japan
  \\
$^4$Universit\'{e} Paris 7, 175 Rue du Chevaleret, 75013 Paris,
France\\
$^5$PRESTO, JST, Kawaguchi, Saitama 332-0012, Japan \\
$^6$Institute for Nano Quantum Information Electronics,\\
The University of Tokyo, Tokyo 113-0033, Japan\\
$^7$Optics Section, Blackett Laboratory \& Institute for
Mathematical Sciences,\\ Imperial College, London SW7 2AZ, United
Kingdom \\
$^8$Department of Physics, Astronomy \& Mathematics, University of
Hertfordshire, Hatfield, AL10 9AB, UK}

\begin{abstract}
We study various distance-like entanglement measures of multipartite states
under certain symmetries. Using group averaging techniques we provide
conditions under which the relative entropy of entanglement, the
geometric measure of entanglement and the logarithmic robustness are
equivalent. We consider important classes of multiparty states, and
in particular show that these measures are equivalent for all stabilizer states, symmetric basis and
antisymmetric basis states. We rigorously prove a conjecture that the closest
product state of permutation symmetric states can always be chosen
to be permutation symmetric. This allows us to calculate the explicit values of
various entanglement measures for symmetric and antisymmetric basis states,
observing that antisymmetric states are generally more entangled. We use these
results to obtain a variety of interesting ensembles of quantum states for which the optimal
LOCC discrimination probability may be explicitly determined and achieved. We
also discuss applications to the construction of optimal entanglement witnesses.
\end{abstract}

\maketitle

\section{Introduction}

The quantification of the entanglement of multipartite quantum states has attracted a great
deal of attention in recent years. Entanglement measures are real valued functions of
quantum states that attempt to quantify the amount of entanglement possessed by different quantum states \cite{VPRK97, relative entropy paper, DHR02, PV07}.
In the case of multipartite entanglement the quantification of entanglement is complicated by the
 fact that multipartite entanglement is known to exist in a variety of different inequivalent forms,
and it is still not clear what the significance of these different forms is \cite{multipartite}.
Nevertheless,
a variety of different entanglement measures have been proposed for the multipartite setting,
with a variety of different motivations \cite{multipartite measure}. Computing these measures, and understanding the relationships between them, is usually very difficult as most measures are defined as the solutions to difficult variational problems.

In this paper we will make progress on this problem by considering
three multipartite entanglement measures, which attempt to quantify
the `distance' between a quantum state and the set of separable
states. The measures that we will consider are the {\it (Global)
Robustness of Entanglement} \cite{robustness paper 2}, the {\it Relative Entropy of
Entanglement} \cite{VPRK97, relative entropy paper}, and the {\it Geometric Measure} \cite{Wie first}.
Although these
quantities do not capture all of the subtleties of entanglement (in
particular the variant of the Geometric measure that we will consider
is not an entanglement monotone on
mixed states, and none of these measures discriminate between the
different forms of multiparty entanglement), all these quantities
have an operational interpretation as bounds on the information that
may be gained by LOCC measurements \cite{Hayashi05}, and the relative entropy of
entanglement in particular has applications to the distillation of
multipartite entanglement \cite{distillation of multipartite}.

The three measures that we consider are related by known
inequalities \cite{Hayashi05}. In this paper we will investigate
conditions under which those inequalities can be shown to be tight.
Our methods rely heavily upon the use of symmetry techniques that
have been applied in papers such as \cite{W89, BBPSSW96, BDSW96, R01, VW01}. We use these methods, together with
some methods from linear algebra, to show that the inequalities in
\cite{Hayashi05} are saturated for stabilizer states (cf.
\cite{MMV06}), antisymmetric states, and symmetric states with fixed
``Hamming weight'' (or fixed ``Type'' for constituent particles of
dimensions greater than a qubit). In the case of the last two
families, explicit expressions may be derived for the entanglement - these
values are summarized in Table \ref{table:one}. Explicit expressions may
also be derived for several families of stabilizer state - we refer the
reader to \cite{MMV06} for details.

In the next section we introduce these measures and the relationship
between them and we discuss the motivation for our investigations in
terms of entanglement witnesses and applications to LOCC information
gain. In section \ref{SEC: Stabilizer and Symm} we present our
general approach and two simple examples (stabilizer states and
symmetric states of fixed {\it type}) where we can use group averaging to prove
equivalence of the measures. For symmetric states of fixed type, we require a useful result
from linear algebra, the Takagi decomposition, which is discussed in Appendix \ref{appendixA}.
In the final section
we apply these methods to certain antisymmetric states, for which
the measures can also be calculated exactly. In Appendix \ref{appendix} we review some
notions of group theory. In Appendix \ref{more general group discussion}, we give a
more general group theoretic
treatment of the problem - the results in the Appendixes \ref{appendix} and \ref{more general group discussion}
are
used throughout the rest of this paper.
\begin{table}[t]
\begin{tabular}{c||l}
{\rm $n$-party state} &\; $E_R$ = $E_g$ = $\log_2(1+R_g)$\\
\hline {\rm Symmetric state } $|S(n,\overrightarrow{k})\rangle$ & $n\log_2 n - \log_2 n! $\\
           &     $- \sum _{j=1}^d (k_j \log_2 k_j - \log_2 k_j! ) $        \\
{\rm Antisymmetric state } $| \Psi_a \rangle$ & $\log_2(n!)$
\end{tabular}
\caption{Summary of the entanglement values obtained for the symmetric (eq.(\ref{snk})) and antisymmetric (eq.(\ref{antisymm})) states that we consider. In addition
the measures are equal for all pure stabilizer states, and equivalent for the (normalised) projector onto
the symmetric subspace. The values of these measures for several classes of stabilizer states are obtained in
\cite{MMV06}}.
\label{table:one}
\end{table}

\section{Outline of problem and motivation}
\label{SEC: motivation}

We now go through the definition of the entanglement measures that we will use throughout this paper, along
with some of their operational interpretations.

 In the following, we assume that our
Hilbert space consists of $m$ local Hilbert space, $\Hi \stackrel{\rm def}{=} \otimes _{i =1}^m \Hi
_i$, each with finite dimensionality. Unless stated otherwise, we treat entanglement and LOCC with
respect to the cut $\otimes _{i=1}^m \Hi _i$, i.e. each Hilbert space is
assumed to belong entirely to distinct parties.

The {\it geometric measure of entanglement} \cite{Wie first, S95, Wie second}, is defined as
\begin{eqnarray} \label{Eqn: Def Eg}
E_g(|\psi\rangle) = \min_{|\Phi\rangle \in {\rm Pro} (\Hi)} -\log_2 (|\langle \Phi | \psi \rangle |^2),
\end{eqnarray}
where ${\rm Pro}(\Hi )$ is the set of product states on $\Hi$. This is the distance between state $|\psi\rangle$ and the
closest product state $|\Phi\rangle$ in terms of fidelity, and has operational significance, for example
in relation to channel capacities \cite{Werner02}. This measure can be extended to the mixed state case
in a natural way via the convex roof method \cite{Wie second}. However, here we will define the ``geometric
measure'' $G(\rho)$ for mixed states as,
\begin{equation}
G(\rho ) \stackrel{\rm def}{=} - \log _2 \max _{\sigma \in {\rm Sep}} \Tr \sigma \rho =
- \log _2 \max _{\ket{\Phi} \in {\rm Pro}(\Hi ) } \bra{\Phi} \rho \ket{\Phi},
\end{equation}
where ``${\rm Sep}$ '' means the set of all separable states on $\Hi = \otimes _{i=1}^m \Hi _i$. Note
that $G(\rho)$ is no longer an entanglement monotone for general mixed states. Indeed, $G(\rho)>0$ for many non-pure
separable states - it can attain its maximal value, for example, on the maximally mixed state. Nevertheless, in the following discussion, $G(\rho)$ works
as a natural extension of $E_g(\ket{\Psi})$ from a mathematical view point, although $G(\rho)$ has
an unusual physical meaning - as it represents a `distance' to the nearest {\it pure} product state (not just
the nearest separable state) it acts more like a measure of both mixedness
and entanglement. In the context of LOCC state discrimination, in which both
 purity and entanglement have operational significance, it is natural that quantities measuring
both entanglement and mixedness should play an important role. $G(\rho)$ is also a useful quantity when constructing entanglement
witnesses as we will see in Sec. \ref{SECsub: Witnesses}.

The {\it relative entropy of entanglement} is defined as the ``distance'' to the closest separable
state with respect to the relative entropy \cite{relative entropy paper},
\begin{eqnarray}
\label{Eqn: DEF rel entropy of entanglement}
    E_R(\rho):=\min_{\omega\in {\rm Sep}} S(\rho||\omega),
\end{eqnarray}
where $S(\rho||\omega) = -S(\rho) - {\rm tr}\{\rho \log_2\omega\}$ is the relative entropy,
$S(\rho)$ is the von Neumann entropy, and ${\rm Sep}$ is the set of separable states. Note that strictly speaking $S(\rho||\omega)$ is not a distance
function. Operationally it tells us, for example, how easy it is to confuse the state $\rho$ for a
separable state in the asymptotic setting \cite{VPRK97}.

The {\it global robustness of entanglement} $R_g(\rho)$ is defined as \cite{robustness paper 2}:
\begin{eqnarray}
\label{Eqn: DEF robustness}
    &R_g(\rho):= \min t \nonumber \\
    &{\rm such ~that~} \exists {\rm ~a ~state~ }\Delta {\rm ,~satisfying} \nonumber \\
    &\omega = {1 \over 1+t }(\rho + t\Delta)\in {\rm Sep},
\end{eqnarray}
where ${\rm Sep}$ is the set of separable states. We can understand this as the minimum (arbitrary) noise
$\Delta$ that we need to add to make the state separable. It can be used also to consider the
robustness of operations against noise \cite{N99}. In the bipartite setting it gives a bound on
how well teleportation can be performed \cite{VV03}. Recently in the general multipartite
setting, it has been shown to be related to optimal entanglement witnesses \cite{B05} (as used in
Sec. \ref{SECsub: Witnesses}). We will often refer to this measure simply as the robustness. For
simplicity in expressions, we will sometimes make use of the logarithmic version, the logarithmic
robustness \cite{logarithmic robustness}:
\begin{eqnarray}
LR_g(\rho):=\log_2(1+R_g(\rho)).
\end{eqnarray}

In a sense, these are very broadly defined measures, and do not pick out many of the possible subtleties
of entanglement in the multipartite scenario (for example the difference between entanglement arising
from multipartite entanglement and that from bipartite entanglement). However, in addition to those
applications already mentioned, they have recently found several
interesting operational interpretations (e.g. \cite{PV07}), including as bounds on how much information can be accessed from states under LOCC \cite{Hayashi05,MMV06}.

In \cite{Hayashi05} (cf. \cite{Wie second}) it has been shown that the following relation holds between the three
different distance-like entanglement measures that we have defined above,
\begin{eqnarray} \label{Eqn: Ent ineq}
r(\rho) \geq E_R(\rho)+ S(\rho) \geq G(\rho),
\end{eqnarray}
where we denote $|A|:=\Tr(A)$, define $P$ as the projector onto the support of $\rho$
\cite{NoteSupport}, and $r(\rho)$
is defined as:
\begin{equation}
r(\rho) := \log_2 |P|\left(1+R(\frac{P}{|P|})\right).
\end{equation}
For pure states the inequalities (\ref{Eqn: Ent ineq})  reduce to
\begin{equation} \label{Eqn: Ent ineq pure}
LR_g(|\psi\rangle) \geq E_R(|\psi\rangle) \geq E_g(|\psi\rangle).
\end{equation}
The difficulty in calculating these measures usually increases from right to left as the defining optimization
problems get harder. We will see that in certain cases we can show equivalence across (\ref{Eqn: Ent ineq}) and (\ref{Eqn: Ent
ineq pure}).

\bigskip

Before we go into any proofs and examples, we will discuss some motivations for studying this problem.
Firstly, showing equivalence across (\ref{Eqn: Ent ineq}) immediately allows the optimization problems
of all measures to be reduced to that of the geometric measure, which is easiest amongst the measures.
This means that all the possible operational interpretations of all the measures can be studied in
terms of the easier, more calculable measure.

In particular we now focus on two applications of these measures - to LOCC state discrimination and the
study of entanglement witnesses.

\subsection{Bounds on state discrimination by separable operations}
\label{SECsub: optimal sep measuremets}

The quantities described in the preceding section all arose naturally
in the authors' previous work \cite{Hayashi05} on LOCC state
discrimination. There it was shown that the measures defined above
supply {\it upper} bounds on the effectiveness of orthogonal state
discrimination when the measurements are implemented separable or
LOCC operations. In this section we will discuss how in situations
of high symmetry the above quantities can also give tight {\it
lower} bounds on what may be achieved by separable operations.

Let us consider at first an ensemble of states $\{p_i,\rho_i\}$ (the $p_i$ are probabilities), the
$\rho_i$ are states that we must discriminate by separable operations. Then we may derive the following
upper bound on the total success probability for discrimination by a separable POVM $\{M_i\}$
\cite{MMV06},
\begin{eqnarray} \label{upper}
P_s = \sum_i p_i \Tr \{M_i \rho_i \} \leq \sum_i p_i  \Tr \{M_i\} 2^{-G(\rho_i)} \nonumber \\
\leq \max_i \{p_i 2^{-G(\rho_i)} \} \sum_i \Tr M_i = D \max_i \{p_i 2^{-G(\rho_i)} \}
\end{eqnarray}
where the first inequality follows from the fact that each $M_i$ is proportional
to a separable state, and $D$ is the total dimension of the system.

This upper bound can be achieved by separable operations in cases where the ensemble is generated by
a local irreducible unitary group acting on some {\it fiducial} state $\phi$, i.e. $\{\rho_i = U_i \phi U^{\dag}_i
|i=1..N \}$, where each state is given to us with uniform prior probability $1/N$. This can be seen as follows.
As all the states are local unitarily equivalent to the fiducial state, the upper
bound becomes :
\begin{eqnarray}
P_s \leq {D \over N} 2^{-G(\phi)}.
\end{eqnarray}
This can be achieved by the separable POVM defined by:
\begin{eqnarray}
M_i := {D \over N} U_i \Omega U^{\dag}_i
\end{eqnarray}
where $\Omega$ is the optimal product state that achieves the geometric measure of the fiducial state
$\phi$.

This motivates the question as to whether `closest separable states' may be used to obtain separable
POVMs that give good lower bounds in other cases. Let us suppose that each state in the
ensemble has a `closest' product state $\psi_i$, i.e. for each $i$ the quantity
\begin{equation}
\rm{tr} \{\rho_i \psi_i\}
\end{equation}
is as large as it can be for an overlap between $\rho_i$ and a separable state. Our goal in trying to find a good separable measurement to discriminate the ensemble will
be to `pretend' that we are instead trying to discriminate these closest
separable states from each other, and use the outcomes to infer information
about the original ensemble $\{\rho_i\}$.
With this goal in mind we write
down the square root measurement for discriminating the ${\psi_i}$:
\begin{equation}
M_i := {{\psi_m}^{-1/2}} p_i \psi_i  {{\psi_m}^{-1/2}} \label{Eqn: square root 1}
\end{equation}
where $\psi_m$ is the mean state:
\begin{equation}
\psi_{mean}:= \psi_m := \sum_i p_i \psi_i.
\end{equation}
For general ensembles $\{\rho_i\}$ with general closest states ${\psi_i}$ there is no guarantee that
the POVM elements $M_i$ defined in equation (\ref{Eqn: square root 1}) will themselves be separable.
However, under the restriction that {\it the mean separable state is itself maximally mixed} then the
$M_i$ defined in equation (\ref{Eqn: square root 1}) will indeed define a separable measurement. In fact
the POVM elements will be given by the separable operators:
\begin{equation}
M_i := p_i D \psi_i
\end{equation}
where $D$ is the total dimension of the system. If we apply this
measurement to the original ensemble, then we find that the optimal probability of successful
discrimination $P_s$ will be bounded by the following expression:
\begin{eqnarray}
P_S \geq &&\sum_i p_i \mbox{tr}\{\rho_i p_i D \psi_i  \} = D \sum_i p^2_i\mbox{tr}\{\rho_i  \psi_i  \} \nonumber \\
= && D\sum_i p^2_i 2^{-G(\rho_i)} \geq D \min_i \{p_i 2^{-G(\rho_i)}\} \label{lower}
\end{eqnarray}
It is not difficult to construct ensembles for which this lower bound matches the upper bound of equation \ref{upper}.
For example, consider any state multi-qubit state $\rho$ for which a closest product state (under the Geometric measure) is an element of the computational basis, such as $|000...\rangle$. Then because the set of product
states $\{X^a \otimes X^b \otimes X^c ...|000...\rangle  | a,b,c...=0,1\}$ define a complete product basis, then the ensemble:
\begin{equation}
\{\{X^a \otimes X^b \otimes X^c ...\rho X^a \otimes X^b \otimes X^c... | a,b,c...=0,1\}\}
\end{equation}
where each state is taken with equal prior probability, will be an example of an ensemble for which
the mean closest product state is maximally mixed. Any such ensemble will also be one for which the upper bound (\ref{upper}) and the lower bound (\ref{lower}) match. Note that this example is not contained
within the examples involving irreducible representations discussed above, as the group $\{X^a \otimes X^b \otimes X^c ...  | a,b,c...=0,1\}$ is not irreducible. Moreover, in such cases the equations (\ref{upper}) and
(\ref{lower}) can be achieved by {\it LOCC} operations, as the POVM defined by the projectors onto the computational
basis may clearly be achieved by LOCC operations. Hence a large number of ensembles may be constructed
for which equations (\ref{upper}) and
(\ref{lower}) provide the exact optimal discrimination probability for both separable and LOCC operations.

The process of constructing such ensembles is by working in reverse - we pick a standard product computational
basis, and then we find states that have these product states as ``closest" separable ones. The ensembles of states
that can be identified in this way are ones for which the lower bounds presented above apply.

Following a similar line of reasoning we may also consider the closest separable states for
the robustness of entanglement. If the states in the ensemble have the closest states
\begin{equation}
\omega_i := {{\rho_i + R_g(\rho_i) \sigma_i} \over {1+R_g(\rho_i)}},
\end{equation}
then as before we may write
down the square root measurement for discriminating the ${\omega_i}$ as:
$M_i := {{\omega_m}^{-1/2}} p_i \omega_i  {{\omega_m}^{-1/2}}$
where $\omega_m$ is the mean state $\omega_{mean}:= \omega_m := \sum_i p_i \omega_i$.
Again, if we assume that $\omega_m$ {\it is itself maximally mixed}, then the
$M_i$ will indeed define a separable measurement. If we apply this
measurement to the original ensemble, then we find that the optimal probability of successful
discrimination $P_s$ will be bounded by the following expression:
\begin{eqnarray}
P_S \geq &&\sum_i p_i \mbox{tr}\{\rho_i p_i D \omega_i  \} = D \sum_i p^2_i\mbox{tr}\{\rho_i  \omega_i  \} \nonumber \\
= && D\sum_i p^2_i\mbox{tr}\{\rho_i  {\rho_i + R_g(\rho_i) \sigma_i \over 1 + R_g(\rho_i)} \} \nonumber \\
\geq && D\sum_i p^2_i {\mbox{tr}\{\rho_i^2 \} \over 1 + R_g(\rho_i)} \nonumber \\
\geq && D \min_i \left( {\mbox{tr}} \{ \rho_i^2 \}{p_i \over 1 + R_g(\rho_i)} \right) \nonumber
\end{eqnarray}
Putting the lower and upper bounds together {\it for ensembles such that the average closest
separable state (for the robustness) is maximally mixed} we find that:
\begin{eqnarray}
D \max_i \{p_i 2^{-G(\rho_i)} \}\geq P_s \geq D \min_i \left( {\mbox{tr}} \{ \rho_i^2 \}{p_i \over 1 + R_g(\rho_i)} \right)\nonumber
\end{eqnarray}
(the upper bound is independent of the nature of the ensemble). We can weaken
the lower bound further by using the inequality {tr$\{\rho_i^2\} \geq 1/|P_i|$}, where
$P_i$ is the projector onto the support of $\rho_i$, in which case the bounds
become:
\begin{eqnarray}
D \max_i \{p_i 2^{-G(\rho_i)} \}\geq P_s \geq D \min_i \left( {p_i \over |P_i|(1 + R(\rho_i))} \right)
\label{lower2}
\end{eqnarray}
As a consequence of (\ref{Eqn: Ent ineq}) one might expect that this lower bound is typically
not as tight as the one derived in equation (\ref{lower}). However, it is quite possible that the requirement that
the mean closest separable state be maximally mixed is not valid for one measure while being valid for
the other, hence the two lower bounds (\ref{lower}),(\ref{lower2}) may separately prove useful in different
cases.

These observations also beg the question as to whether the stringent constraint on the nature of the ensemble -
the lower bounds are only valid when the mean closest separable states is maximally mixed - may
be relaxed. Some generalisations should be possible - for instance, if the average mean state is sufficiently
close to maximally mixed, then a perturbation of the above approach should lead to similar
 bounds as all the quantities considered above are continuous. However, it would be of more general interest to consider how one can define a separable analogue of the square root measurement in situations where $\omega_m$ is not constrained at all. A more general approach, for example,
would be to write the global square root measurement, and compute
bounds on the minimal noise required to make that global POVM separable. We will not, however, pursue this approach any further here, as we hope to pursue it in future work.

\subsection{Optimal Entanglement Witnesses}
\label{SECsub: Witnesses}

We will now see how two of the entanglement measures considered, the
robustness and the geometric measure, are naturally related to the
concept of entanglement witnesses. The geometric measure can be used
to define a particular entanglement witness which we will denote
$W_G$. The robustness of entanglement can be considered as a
quantification of the amount a state violates a kind of optimal
witness which we denote $W_R$. As we shall see, if the geometric measure
and logarithmic robustness are equal, then both $W_G$ and $W_R$ are optimal in
the sense of $\rho$ optimality considered in \cite{B05}. Note that this
notion of optimality is actually different to the notions of optimality considered
both in \cite{T00} and \cite{LKCH00} - in those papers a witness is only said to be
optimal if it is impossible to find another witness that detects a strictly larger set of
entangled states. The notion of $\rho$ optimality is likely to be more relevant when
considering the statistical significance of violations in
experimental implementations.

An entanglement witness $W$ is a Hermitian operator (hence an observable) such that for all separable
states $\omega$, $\Tr(\omega W)\geq 0$, and for some entangled state $\rho$, $\Tr(\rho W)<0$. $W$ is
said to witness the entanglement of $\rho$ \cite{T00}.

Similar to those used in, for example Ref. \cite{BEKGWGHBLS04}, it can easily be seen that the geometric
measure $G(\rho)$ naturally defines a normalised entanglement witness associated to state $\rho$,
\begin{eqnarray}
W_G(\rho) :&=& {1 \over \alpha }\left( \alpha \one - \rho \right) \nonumber \\
\alpha &=& \max_{\omega \in SEP} \Tr(\rho \omega) = 2^{-G(\rho)}.
\end{eqnarray}

Some of these witnesses may be trivial, because if the maximal eigenvalue of $\rho$
corresponds to a product eigenstate, then the witness will not detect any entangled states
at all. However, if the maximal eigenvalue of $\rho$ is non-degenerate and corresponds to an entangled eigenstate,
then the witness will certainly detect some entangled states.

A so-called {\it $\rho$-optimal} entanglement witness ($\rho$-OEW)
relative to a set $\mathcal{M}$ is a witness $W^{\mathcal{M}}_\rho$,  that is associated to a state $\rho$, and which
satisfies \cite{T02}
\begin{eqnarray}
\Tr(W^{\mathcal{M}}_\rho \rho) =\min_{W\in\mathcal{M}}\Tr(W\rho),
\end{eqnarray}
where $\mathcal{M}$ is a compact subset of entanglement witnesses \cite{NoterhoOEW}. In this way a $\rho$-OEW
is one which is violated maximally for the state $\rho$ at hand, for a given class of witnesses
$\mathcal{M}$. Experimentally we may like to choose such a witness since the violation would then be
the most visible.

We will see that equality of the logarithmic robustness and the
geometric measure implies that the witnesses $W_G$ are
$\rho$ optimal for the set ${\mathcal{M}}$ of entanglement witnesses
satisfying $\mathcal{M}=\{W|W\in\mathcal{W}, W\leq \one
\}$. This is the set of witnesses that can be associated
in a special way to the robustness of
entanglement: In \cite{B05} it is shown that the robustness is given by
\begin{eqnarray}
R_g(\rho) =\max \{0,-\min_{W\in \mathcal{M}}\Tr(W\rho)\}, \label{eqn: Robustness is opt witness}
\end{eqnarray}
where $\mathcal{M}=\{W|W\in\mathcal{W}, W\leq \one
\}$.  This implies that, for any state $\rho$, if there exists a
witness, we write $W_R$ such that $R_g(\rho)=-\Tr(\rho W_R)$, then $W_R$ is
$\rho$-OEW relative to the set $\mathcal{M}=\{W|W\in\mathcal{W}, W\leq \one
\}$.
\bigskip

\noindent{\bf Proposition}: For a projection state $\rho=\frac{P}{|P|}$, if we have equivalence of
measures $\log_2(|P|(1+R_g(\rho)))=E_R(\rho)+S(\rho) = G(\rho)$, then the normalised witness
$W_G(\rho)$ is a $\rho$-OEW relative to the set $\mathcal{M}=\{W|W\in\mathcal{W}, W\leq \one
\}$.

\bigskip

\noindent {\bf Proof}: If $G(\rho)=\log_2 \left( |P|(1+R_g(\rho))
\right)$, then
\begin{eqnarray}
R_g(\rho) = \frac{2^{G(\rho)}}{|P|} - 1.
\end{eqnarray}
The proposition is proved by comparing this to the expectation value
of $W_G(\rho)$ for $\rho$:
\begin{eqnarray}
-\Tr(W_G(\rho) \rho ) &=& - 1 +  \frac{2^{G(\rho)}}{|P|} = R_g(\rho).
\end{eqnarray}
By (\ref{eqn: Robustness is opt witness}),
$W_G(\rho)$ is also a $\rho$-OEW relative to the set $\mathcal{M}=\{W|W\in\mathcal{W}, W\leq \one
\}$.  $\square$

\section{Outline of approach: Stabilizer states and Permutation invariant basis states}
\label{SEC: Stabilizer and Symm}

The essence of the argument to prove equivalence of the measures
across (\ref{Eqn: Ent ineq}) is to take the product state
$|\Phi\rangle$ which achieves the geometric measure (\ref{Eqn: Def
Eg}), and perform a local ``twirling'' operation (a group
averaging), to give a separable mixed state. If the symmetries have
a suitable structure, or if the product state $|\Phi\rangle$ has
certain properties, then the twirled version of $|\Phi\rangle$ can
be a good candidate for the state $\omega$ in the optimisation for
the global robustness (\ref{Eqn: DEF robustness}). This then gives
an upper  bound to the robustness which which sits on the left of
(\ref{Eqn: Ent ineq}), (\ref{Eqn: Ent ineq pure}). We will see that
for certain states this  upper  bound matches the geometric measure,
hence implying equality across (\ref{Eqn: Ent ineq}), (\ref{Eqn: Ent
ineq pure}). For this to work it is essential that the twirled
product state be of the correct form (\ref{Eqn: DEF robustness}). A
more formal group theoretical statement of this is given in appendix
\ref{appendix}. In general these conditions must be checked by
knowing the closest product state $|\Phi\rangle$ (see Theorem $1$
for projection states and Theorem $2$ for pure states in appendix
\ref{appendix}). In certain cases some group symmetry properties of
$|\Phi\rangle$ will suffice. This is the case for the symmetric
bases states as we will see. In other cases the conditions may be
satisfied simply by the properties of the group averaging and we do
not need to know anything about the state $|\Phi\rangle$ (see
Theorem 3 for projection states and Theorem 4 for pure states in appendix
\ref{appendix}). This is the case for the stabilizer states as we
will see. In this section we will first give a sketch of the ideas,
and two sets of examples which illustrate the methods that we will
use.

If $|\Phi\rangle$ is the closest product state to pure state $|\psi\rangle$, the effect of
averaging over some group $\{ \rm U \}$ is essentially to project onto the invariant subspaces (see Lemma \ref{definition
of representation})
\begin{eqnarray}
\omega' = \int {\rm U} |\Phi\rangle\langle \Phi| {\rm U }^{\dag}d{\rm U} = \sum_i P_i |\Phi\rangle\langle \Phi| P_i,
\end{eqnarray}
where $P_i$ are the projectors onto the invariant subspaces. Since the ${\rm U}$ are local, $\omega'$ is
separable. In order to be a valid candidate for the robustness state $\omega$ in (\ref{Eqn: DEF robustness}),
we require that it is possible to reach $\omega'$ by adding noise to $|\psi\rangle \langle \psi|$. This is certainly possible
if for some $i$ we have \\

(i) $P_i |\Phi\rangle \langle \Phi| P_i = \lambda |\psi\rangle \langle \psi|$,\\

\noindent hence if $|\psi\rangle$ is invariant under the action of the group. Further, if we also have \\

(ii) $\lambda = 2^{-E_g(|\psi\rangle)}$,\\

\noindent then it can be shown quite easily that
$LG(|\psi\rangle)=E_g(|\psi\rangle)$, hence we have equality across
(\ref{Eqn: Ent ineq pure}) (see Theorem 1 and Theorem 2 in appendix
\ref{appendix} for a more general group theoretic statement of this
fact).

Both (i) and (ii) are immediately satisfied if $|\psi\rangle$ is
itself a full  invariant subspace, i.e. one of the $P_i$ is itself
the projector $|\psi\rangle\langle\psi|$ (see Theorem 3 and Theorem
4 in appendix \ref{appendix}). This will be the case for our first set of
examples below, the stabilizer states. If this is not the case, we
need to find other ways to check that (i) and (ii) are explicitly
satisfied (note, $|\psi\rangle$ must still be invariant). We do
this by explicitly finding the closest product state and checking.
The symmetric basis states provide an example of this case, as we
will see below.

\subsection{Stabilizer States}
\label{SECsub: Stabilizer}

A stabiliser states $|S\rangle$ is defined by the associated group $S=\{G_i\}_{i=1}^{2^n}$, where $G_i$
are made up of local Pauli operators, which stabilize the state in the sense of the eigen-equations
\cite{G99}
\begin{eqnarray} \label{eqn: def stabilizer}
G_i|S\rangle = |S\rangle, ~~ \forall G_i \in S.
\end{eqnarray}
The group $S$ is called the {\it stabilizer group}, and the equations (\ref{eqn: def stabilizer})
completely characterize the state. In fact, by considering the plus and minus mutual eigen-states of
$S$ we define a complete basis. Taking any $n$ generators, we define the $2^n$ basis states
$\{|S_{g_1,g_2..g_n}\rangle\}$, with
\begin{eqnarray}
G_i|S_{g_1,g_2..g_n}\rangle =(-1)^{g_i} |S_{g_1,g_2..g_n}\rangle
\end{eqnarray}
where $g_i=1,0$ corresponding to eigen values $+1$ or $-1$
respectively, label the basis states.  These states are exactly the
invariant subspaces of the stabilizer group, i.e.
$P_{\bar{g}}=|S_{\bar{g}}\rangle \langle S_{\bar{g}}|$, where
$\bar{g}$ is the binary list $g_1,g_2..g_n$. The stabilizer state
(\ref{eqn: def stabilizer}) is then $|S\rangle=|S_{0,0,..0}\rangle$.

Denoting $|\Phi_S\rangle$ as the closest product state, we have
\begin{eqnarray}
E_g(|S\rangle) = -\log_2 |\langle \Phi_S | S\rangle|^2.
\end{eqnarray}

We construct our candidate for the closest separable state $\omega$ in \ref{Eqn: DEF robustness}, by
averaging (or ``twirling'') over a local group, in this case, the stabilizer group. We thus define
\begin{eqnarray}\label{eqn: twirled stab prod}
\omega ' &=& \sum_{G_i\in S} G_i |\Phi_S\rangle \langle \Phi_S| G_i \nonumber \\
&=& \sum_{\bar{g}}|\langle S_{\bar{g}} |\Phi_S \rangle|^2 |S_{\bar{g}} \rangle \langle S_{\bar{g}} |.
\end{eqnarray}

Since the operators $G_i$ are local, the state $\omega'$ is a
separable state, we can hence consider it as a candidate for closest
separable state. For any candidate state $\omega' =
\frac{1}{1+t'}(\rho + t'\delta)$, we have that $t'\geq R_g(\rho)$.
State (\ref{eqn: twirled stab prod}) is of this form for $|S\rangle$
with $t'=\frac{1}{|\langle S |\Phi_S \rangle|^2}-1 =
2^{E_g(|S\rangle)}-1$. Hence we have
\begin{eqnarray}
E_g(|S\rangle)  \geq \log_2(1+R_g(|S\rangle)) \geq E_R(|S\rangle) \geq E_g(|S\rangle),
\end{eqnarray}
proving equality across all measures, i.e.
\begin{eqnarray}
\log_2(1+R_g(|S\rangle)) = E_R(|S\rangle) = E_g(|S\rangle).
\end{eqnarray}

\bigskip

We can now consider what this means in terms of measurements and
witnesses from our earlier discussion.
Suppose that we are working in a basis where
closest product state to $|S_{0,0,0..}\rangle$ is $|000...\rangle$, then the optimal
probability of discriminating the ensemble of graph states
$\{X^a \otimes X^b \otimes X^c ...|S_{0,0,0..}\rangle\| a,b,c...=0,1\}$
(all with equal a-priori probability $p=2^{-n}$), is given exactly by equation
(\ref{lower}) - this follows from the discussion in section \ref{SEC: motivation}.
Using the explicit formulae presented in \cite{MMV06} for the entanglement of
a variety of classes of stabilizer state, many ensembles
of graph states may be constructed whose optimal LOCC discrimination
probability may be obtained in this way.

 To define the proposed entanglement witness $W_G$ we
need the value of the geometric measure. Here we do not have it,
however, we can say that for any cases where it is known such
witness will also hold as $W_R$. Examples of where it is known for
many important stabilizer states including cluster states is given
in \cite{MMV06}.

\subsection{Permutation Symmetric States}
\label{SECsub: Symmetric}

In the previous case the state itself is an invariant subspace of
the group, and this is sufficient for showing the equivalence of the
measures (as stated more precisely in Theorem 3 and 4 of Appendix
\ref{appendix}). If we also know the state $|\Phi\rangle$ which
gives the geometric measure we can relax this requirement a little
(Theorems 1 and 2 in Appendix \ref{appendix}). We will do just that to prove
equivalence of these measures for the so called symmetric basis
states.

In $\Hi = (\mathbb{C}^d)^{\otimes n}$, symmetric basis states
$\ket{S(n,\overrightarrow{k})}$, which form a basis of the symmetric
subspace $\mathcal{S}_n$, are defined as
\begin{widetext}
\begin{equation}
\ket{S(n,\overrightarrow{k})} :=\frac{1}{\sqrt{C_{n,\overrightarrow{k}}} }\sum _{\overrightarrow{i}
\in {\rm perm} }\ket{\overbrace{0 \cdots 0}^{k_0} \overbrace{1 \cdots 1}^{k_1} \cdots \overbrace{d-1
\cdots d-1}^{k_{d-1}}}, \label{snk}
\end{equation}
\end{widetext}
where the summation is over all permutations of the sequence $(\overbrace{0 \cdots 0}^{k_0} \cdots
\overbrace{d-1 \cdots d-1}^{k_{d-1}} ), $  (that is, a $n$-length sequence in which ``$i$'' appears
just $k_i$ times ), and $C_{n,\overrightarrow{k}} := |\rm perm | = \frac{n!}{\Pi _{j=1}^d k_j !}$. We
also note that $\overrightarrow{k} = (k_0, \cdots,  k_{d-1} )$ satisfies $\sum _{a=0}^{d-1} k_a = n $.

For symmetric basis states, the value of the geometric measure of entanglement is already known
\cite{Wie second}, under the assumption that the closest product state is also symmetric,
\begin{equation} \label{Eqn: Eg snk}
E_g(\ket{S(n,\overrightarrow{k})})  = n\log_2 n - \log_2 n! - \sum
_{j=1}^d (k_j \log_2 k_j - \log_2 k_j! ),
\end{equation}
and a closest product state is given by
\begin{eqnarray}\label{Eqn: closest prod SNK}
|\Phi\rangle=\left(\sum _{l=1}^d
\sqrt{\frac{k_l}{n}}\ket{l}\right)^{\otimes n}.
\end{eqnarray}

Before we show the equivalence of the entanglement measures, we will first prove rigourously
the working assumption leading to (\ref{Eqn: Eg snk}), (\ref{Eqn: closest prod SNK}), by using symmetry
arguments.

\begin{Lemma}\label{geometric measure symmetric}
If $\ket{\Psi } \in \mathcal{S} _n$, then, there exist a closest
product state $|\Phi\rangle$ in the symmetric Hilbert space, thus,
\begin{equation}\label{Eq geometric measure symmetric}
|\Phi\rangle = |\phi\rangle^{\otimes n}, ~ ~ E_g(\ket{\Psi})= -\log _2 \max _{\ket{\phi}\in \Hi }
|\bra{\phi}^{\otimes n}\ket{\Psi}|^2,
\end{equation}
where $\mathcal{S} _n$ is symmetric subspace of $\Hi^{\otimes n}$.
\end{Lemma}

\noindent {\bf Proof of Lemma 1:} We prove this in \ref{SEC: App symm}.

 Using this lemma it is possible to show that (\ref{Eqn: closest prod SNK}) gives the closest product state
\cite{Wie first}. For completeness we give a simplified proof of this:

\begin{Lemma}
If $\ket{\Psi } = \ket{S(n,\overrightarrow{k})}$, then a closest product state $|\Phi\rangle$ for the geometric
measure is given by Eq. (\ref{Eqn: closest prod SNK}), i.e.:
\begin{eqnarray}
|\Phi\rangle=\left(\sum _{l=1}^d
\sqrt{\frac{k_l}{n}}\ket{l}\right)^{\otimes n}.
\end{eqnarray}
\end{Lemma}

\begin{Proof} By Lemma \ref{geometric measure symmetric}, $|\bra{S(n, \overrightarrow{k}} \ket{\Phi}|$ attains its
maximum when $\ket{\Phi}$ can be written as $\ket{\Phi} = \ket{\phi}^{\otimes n}$ for a local
state $\ket{\phi} \in \Hi$. Moreover, since all coefficients of $\ket{S(n, \overrightarrow{k})}$ are
positive in the computational basis, $|\bra{S(n, \overrightarrow{k})} \ket{\phi }^{\otimes n}|$ attains its
maximum when all coefficients of $\ket{\phi}$ are positive in the computational basis. Thus, we can write
down $\ket{\phi}$ as $\ket{\phi} = \ket{\overrightarrow{p}} \stackrel{\rm def}{=} \sum _{l=1}^d
\sqrt{p_l}\ket{l}$ for some probability distribution $\overrightarrow{p}$. Using this we can derive an upper
bound as follows,
\begin{eqnarray}
\bra{S(n, \overrightarrow{k})} \ket{\overrightarrow{p} }^{\otimes n} &=&
\sqrt{C_{n,\overrightarrow{k}}} \Pi _{l=1}^d \sqrt{p_l}^{k_l} \nonumber\\
&=& \sqrt{C_{n,\overrightarrow{k}}} 2^{\frac{n }{2 }(\sum_{l=1}^d
\frac{\overrightarrow{k} }{n}\log_2 p_l)} \nonumber\\
&=& \sqrt{C_{n,\overrightarrow{k}}}
2^{\frac{n}{2}H(\frac{\overrightarrow{k}}{n})-D(\frac{\overrightarrow{k}}{n}\|\overrightarrow{p})}
\nonumber\\
&\le & \sqrt{C_{n,\overrightarrow{k}}} 2^{\frac{n}{2} H(\frac{\overrightarrow{k}}{n})},
\label{inequality of relative entropy}
\end{eqnarray}
where $H(\overrightarrow{p})$ is the Shannon entropy, $D(\overrightarrow{p}\| \overrightarrow{q})$ is the
Classical relative entropy, and the inequality follows from the positivity of the relative entropy. In 
(\ref{inequality of relative entropy}), equality holds if and only if
$\overrightarrow{p}= \overrightarrow{k}/n$, since a necessary and sufficient condition for
$D(\overrightarrow{p}\|\overrightarrow{q}) = 0$ is $\overrightarrow{p} = \overrightarrow{q}$. \hfill
$\square$
\end{Proof}

\noindent We are now ready to show equality of the measures
\begin{eqnarray}\label{robustness symmetric}
&\quad & \log _2 (1+R_g(\ket{S(n, \overrightarrow{k} ) }) \nonumber \\
&=& E_r(\ket{S(n, \overrightarrow{k} )})=
E_g(\ket{S(n, \overrightarrow{k})}) \nonumber \\
&=& n\log_2 n - \log_2 n! - \sum _{j=1}^d (k_j \log_2 k_j - \log_2 k_j! ).
\end{eqnarray}
To show this we average over the group ${\rm U}(1)\times \cdots \times {\rm U}(1)$, with
representation
\begin{widetext}
\begin{eqnarray}
{\rm U}(\theta_1, \theta_2..\theta_{d-1})
 =  (\sum _{j_1 = 0}^{d-1}
\exp(i\theta _{j_1 })\ket{j_1}\bra{j_1})\otimes \cdots
 \otimes (\sum
_{j_n = 0}^{d-1} \exp(i\theta _{j_n}
)\ket{j_n}\bra{j_n})\label{Eqn: sym state local u}
\end{eqnarray}
\end{widetext}
The symmetric states $\ket{S(n,\overrightarrow{k})}\bra{S(n,\overrightarrow{k})}$ are invariant elements of
this representation if we choose $\theta _0 = 0$. However, they are not the total invariant
subspaces.

At this point to check that the twirled states are of the correct
form we could simply apply (\ref{Eqn: sym state local u}) with
$\theta=0$ to the state (\ref{Eqn: closest prod SNK}) and it easily
follows that
\begin{widetext}
\begin{eqnarray}
\omega '
&=&  \int _0^{2\pi} \cdots \int _0^{2\pi} {\rm U}(\theta_1,
\theta_2..\theta_{d-1}) \ket{\Phi}\bra{\Phi
} {\rm U}(\theta_1, \theta_2..\theta_{d-1}) ^\dag d\theta _1 \cdots d\theta _{d-1}  \nonumber \\
&=& \sum _{\overrightarrow{k}} \left| \bra{\Phi}
\ket{S(n,\overrightarrow{k})}\right| ^2
\ket{S(n,\overrightarrow{k})}\bra{S(n,\overrightarrow{k})}, \label{Eqn:
proof of symmetric basis state i}
\end{eqnarray}
\end{widetext}
which by construction is separable and is of the appropriate form
and hence proves equality of the measures.

In fact, however, this can also be seen without knowing the exact
state $|\Phi\rangle$ itself, but using only the fact that it must be
symmetric (Lemma 1).

It can easily be seen that the invariant subspace of this unitary
group consists of the subspace of the fixed ``Type'' (or fixed
"Hamming weight") $\mathcal{A}_{\overrightarrow{k}}$; by means of
the $d$-dimensional vector $\overrightarrow{k} = (k_0,k_1, \cdots,
k_{d-1})$ satisfying $k_i \ge 0$ and $\sum _{i=0}^{d-1} k_i =n$, the
subspace $\mathcal{A}_{\overrightarrow{k}}$ is defined as
$\mathcal{A}_{\overrightarrow{k}} = {\it span } \{ \ket{a}\bra{b} \
| \ a, b \in {\rm Type}(\overrightarrow{k})   \}$, where ${\rm
Type}(\overrightarrow{k})$ is the set of sequences derived by
permutations of $\{\overbrace{0, \cdots, 0}^{k_0}, \overbrace{1,
\cdots, 1}^{k_1}, \cdots ,\overbrace{d-1, \cdots, d-1}^{k_{d-1}} \}$
(sequences of ``Type $\overrightarrow{k}$'' ). Thus, the
projection operator corresponding to the total invariant subspaces
$\mathcal{A}_{\overrightarrow{k}}$ can be written down as
$P_{\mathcal{A}_{\overrightarrow{k}}} = \sum _{a \in {\rm
Type}(\overrightarrow{k})}\ket{a}\bra{a}$. We thus need to check that the
twirled $|\Phi\rangle$ is of the correct form.

We choose a closest product state from the symmetric Hilbert space (Lemma \ref{geometric measure symmetric}),
and average (\ref{Eqn: closest prod SNK}) over ${\rm U}(\theta_1, \theta_2..\theta_{d-1})$ to get
\begin{widetext}
\begin{eqnarray}
\omega ' = && \int _0^{2\pi} \cdots \int _0^{2\pi} {\rm U}(\theta_1, \theta_2..\theta_{d-1})
\ket{\Phi}\bra{\Phi
} {\rm U}(\theta_1, \theta_2..\theta_{d-1}) ^\dag d\theta _1 \cdots d\theta _{d-1}  \nonumber \\
&&\quad= \sum _{\overrightarrow{k}}P_{\mathcal{A}_{\overrightarrow{k}}}\ket{\Phi}\bra{\Phi}P_{\mathcal{A}_{\overrightarrow{k}}} \nonumber\\
&&\quad= \sum _{\overrightarrow{k}}P_{\mathcal{A}_{\overrightarrow{k}}}(\sum _{\overrightarrow{l}} \ket{S(n,\overrightarrow{l})}\bra{S(n,\overrightarrow{l})})\ket{\Phi}
\bra{\Phi}(\sum _{\overrightarrow{m}} \ket{S(n,\overrightarrow{m})}\bra{S(n,\overrightarrow{m})})P_{\mathcal{A}_{\overrightarrow{k}}} \nonumber\\
&&\quad= \sum _{\overrightarrow{k},\overrightarrow{l},\overrightarrow{m}}\delta_{\overrightarrow{k}\overrightarrow{l}}\delta_{\overrightarrow{k}\overrightarrow{m}} \langle S(n,\overrightarrow{l}) | \Phi  \rangle
\langle \Phi | S(n,\overrightarrow{m})  \rangle  \ket{S(n,\overrightarrow{l})}\bra{S(n,\overrightarrow{m})} \nonumber \\
&&\quad= \sum _{\overrightarrow{k}} \left| \bra{\Phi}
\ket{S(n,\overrightarrow{k})}\right| ^2
\ket{S(n,\overrightarrow{k})}\bra{S(n,\overrightarrow{k})}, \label{Eqn: proof of symmetric basis state ii}
\end{eqnarray}
\end{widetext}
where we use Lemma \ref{invariant element in invariant subspace} in appendix
\ref{appendix} in the second part, the fact that a closest product
state $\ket{\Phi}$ is in the symmetric Hilbert space and the
equation $\left ( \sum _{\overrightarrow{k}}
\ket{S(n,\overrightarrow{k})}\bra{S(n,\overrightarrow{k})} \right )
\ket{\Phi}=\ket{\Phi} $ in the third part, and the fact $
P_{\mathcal{A}_{\overrightarrow{k}}} \ket{S(n,\overrightarrow{l})} =
\delta_{\overrightarrow{k}\overrightarrow{l}}\ket{S(n,\overrightarrow{l})}$
in the fourth part. Since the original state $|\Phi\rangle$ is
separable, and only local unitaries are used, the final state
$\omega'$ is separable. We see that the state $\omega '$ is now a
candidate state for the closest separable state for the robustness,
and we again get equivalence of the measures Eq.(\ref{robustness
symmetric}) in the same way as the stabilizer states. Note that, in
comparison with the case of the stabilizer states, we must use
additional information about the nearest product state $\ket{\Phi}$
in the proof of Eq.(\ref{robustness symmetric}); that is, in Eq.
(\ref{Eqn: proof of symmetric basis state ii}), we use the fact that a
closest product state can be chosen from the symmetric Hilbert space
(Lemma \ref{geometric measure symmetric}). This shows that we
generally cannot conclude the equivalence of the entanglement
measures only by invariance of a state under local unitary group
actions (see Theorem \ref{group representation extra 2} in appendix
\ref{appendix}).

\bigskip

We now turn again to the topics of separable measurements and
witnesses. Again the methods of Sec. \ref{SECsub: optimal
sep measuremets} can be applied to obtain ensembles
of states that are local unitarily equivalent to the
symmetric basis states, and for which the optimal
LOCC discrimination procedure is given by a simple
product measurement. We may also easily apply
the discussion concerning optimal entanglement witnesses.
Since in this case we know the value of $E_G$ we can define
the entanglement witness as in section \ref{SECsub: Witnesses},
\begin{eqnarray}
W_G(|S(n,k)\rangle)&=& \frac{1}{\alpha}(\alpha \one - |S(n,k)\rangle \langle
S(n,k)|)\nonumber \\
\alpha &=&
C_{n,\overrightarrow{k}}\prod_{l=1}^d\left(\frac{k_l}{n}\right)^{k_l}
\end{eqnarray}
which by the equality of the measures will be $\rho$-OEW.

\section{Further Examples: Multi-partite states related to the tensor product
representation of ${\rm U}(n)$} \label{SEC: Further examples}

We now consider a set of further examples. Suppose our Hilbert space
is $\Hi = (\mathbb{C}^d) ^{\otimes n}$. We consider the tensor
product representation of ${\rm U}(d)$, that is, $\pi : {\rm U} \in {\rm U}(d) \mapsto
\overbrace{{\rm U} \otimes \cdots \otimes {\rm U}}^n \in \B _2(\Hi)$. This
representation clearly only involves local unitary operations. It
is well known that, by means of ``{\it Weyl's unitary trick}'',
there exists a natural bijection between all irreducible
representations derived from the above representation of ${\rm U}(d)$ and
all irreducible representations which are derived from the tensor
product representation of $GL(d)$, that is, $A \in GL(d)  \mapsto
\overbrace{A \otimes \cdots \otimes A}^n \in \B _2(\Hi)$ \cite{weyl,
goodman-wallach, Iwanami}. Moreover, by ``{\it Schur duality}'', the
tensor product representation of $GL(d)$ can be decomposed as
follows\cite{weyl, goodman-wallach},
\begin{equation}
(\mathbb{C}^d) ^{\otimes n} \cong \bigoplus _{\lambda \in {\rm
Par} (n,d)} \ G^{\lambda } \otimes F_d^{\lambda},
\end{equation}
where  ${\rm Par} (n,d)$ is a partition of $n$ with depth $d \le
n$, that is, a set of $\lambda \in \mathbb{N}^d$ satisfying
$\lambda _1 \ge \cdots \ge \lambda _d$ and $\sum _{i=1}^d \lambda
_i = n$, $G^{\lambda}$ is the space of an irreducible
representation of the symmetric group of degree $n$ (which we denote
$\mathfrak{G}_n$) defined by partition $\lambda $, and
$F_d^{\lambda}$ is the representation space of the irreducible
representation of $GL(d)$ with the highest weight $\lambda $
\cite{weyl, goodman-wallach}. Using Young tableaux terminology,
$\lambda \in {\rm Par}(n,d)$ corresponds to a Young tableau which
has $\lambda _k$ boxes in the $k$th row. Since this representation
$(\pi, \Hi) $ can be decomposed by
\begin{equation}
(\pi, \Hi) \cong (\bigoplus _{\lambda \in {\rm Par} (n,d)} \
G^{\lambda } \otimes F_d^{\lambda}, \bigoplus _{\lambda \in {\rm
Par} (n,d)} \ I_{G^{\lambda} } \otimes \pi _{\lambda}),
\end{equation}
where $\pi _{\lambda}$ is an irreducible representation with
highest weight $\lambda$, we can apply Theorem \ref{group
representation main} for this representation of ${\rm U}(n)$. In order
to apply Theorem \ref{group representation main} for the
projection states corresponding to subspace $F_d^{\lambda}$, the
dimension of $G_{\lambda}$ must be one. Since the
dimension of $G _{\lambda}$ is given by the number of {\it
standard Young tableaux} (that is, a Young tableau in which the
numbers form an increasing sequence along each line and along each
column) corresponding to the partition $\lambda$, a necessary and
sufficient condition for $\dim G_{\lambda} =1 $ is $\lambda = (n,
0, \cdots, 0), {\rm or} \ ( \overbrace{1, \cdots, 1}^{n} )$. It
is also well known that the representation space $F_d^{\lambda}$
of partition $\lambda = (n, 0, \cdots, 0)$ corresponds to the
symmetric Hilbert space $\mathcal{S}_n $, and the representation
space of partition $\lambda = (1, \cdots, 1)$ corresponds to
the anti-symmetric Hilbert space $\mathcal{A}_n$, which only exists
under the condition $n \le d$. Hence we have proven the
following Corollary,
\begin{Corollary} \label{group representation symmetric and anti-symmetric}
In $\Hi = (\mathbb{C}^d)^{\otimes n}$, the projection states
$\frac{P}{\Tr P}$ corresponding to the symmetric (${\rm Ran}P =
\mathcal{S}_n$) and anti-symmetric (${\rm Ran}P = \mathcal{A}_n$)
Hilbert spaces satisfy
\begin{equation}
\log _2 (1+R_g(\frac{P}{\Tr P})) = E_R(\frac{P}{\Tr P}) =
G(\frac{P}{\Tr P})-\log_2 \Tr P.
\end{equation}
\end{Corollary}
As we will see in the following part, an anti-symmetric basis
state is an example to which this corollary may be applied.

{\it Anti-symmetric basis states.}
Suppose $\Hi = (\mathbb{C}^n)^{\otimes n}$, $n \le d$, and
$\ket{\Psi _a} \stackrel{\rm def}{=} \ket{1} \wedge \cdots \wedge
\ket{n}$, (we call $\ket{\Psi _a}$ an anti-symmetric basis
state), where $\{ \ket{i} \}_{i=1}^n $ is an orthonormal basis of
$\mathbb{C}^n$, and $\wedge$ is the wedge product ($\ket{a} \wedge
\ket{b} = \frac{1}{\sqrt{2}}(\ket{a} \otimes \ket{b} - \ket{b}
\otimes \ket{a})$). Since for the irreducible
representation $(\pi_{(1, \cdots, 1)}, F_d^{(1, \cdots, 1)})$,
$F_d^{(1, \cdots, 1)} = \mathcal{A}_d = \mathbb{C} \ket{\Psi _a}$,
by means of Theorem \ref{group representation symmetric and
anti-symmetric}, we have equivalence of distance like measures
\begin{equation} \label{equality for anti-symmetric}
\log _2 (1+R_g(\ket{\Psi _a})) = E_R(\ket{\Psi _a}) =
E_g(\ket{\Psi _a}).
\end{equation}
Moreover, the value of the geometric measure of entanglement is known
in this case as follows \cite{Bravyi}:
\begin{Lemma}\label{geometric measure of anti-symmetric basis states}
In $\Hi = (\mathbb{C}^d)^{\otimes n}$, anti-symmetric basis states
\begin{eqnarray}
\ket{\Psi _a} &\stackrel{\rm def}{=}& \ket{l} \wedge \cdots \wedge \ket{l+n} \nonumber \\
&=& \frac{1}{N!} \sum_{\{k_l \}}\epsilon_{k_1,k_2,..,k_N} | \alpha_{k_1},..,\alpha_{k_N} \rangle \label{antisymm}
\end{eqnarray}
satisfy
\begin{equation}\label{Eq geometric measure of anti-symmetric basis states}
E_g(\ket{\Psi _a}) = \log_2 n!,
\end{equation}
where $\epsilon_{k_1,k_2,..,k_N}$ is the Levi-Civita symbol, $n \le d$, $1 \le l \le d-n$, and $\{
\ket{i} \}_{i=1}^d$ is an orthonormal basis on $\mathbb{C}^d$.
\end{Lemma}

\begin{Proof}
Firstly, the entanglement of $\ket{l} \wedge \cdots \wedge \ket{l+n}$ in
$(\mathbb{C}^d)^{\otimes n}$, and the entanglement of $\ket{1} \wedge \cdots \wedge \ket{n}$ in
$(\mathbb{C}^n)^{\otimes n}$ are equivalent, because they can be interconverted by LOCC. Thus, we
only consider the case $\ket{\Psi _a} = \ket{1} \wedge \cdots \wedge \ket{n}$. Therefore, all we have
to do is to calculate the value of the geometric measure of entanglement for $\ket{1} \wedge \cdots \wedge
\ket{n}$.
From the definition of the wedge product we can easily see that
\begin{eqnarray*}
&\quad & \bra{\phi _1} \otimes \bra{\phi _1} \otimes \bra{\phi _3 } \otimes \cdots \otimes \bra{\phi _n}
\ket{\Psi _a} \\
&=& \bra{\phi _1} \otimes \bra{\phi _1} \otimes \bra{\phi _3 } \otimes \cdots \otimes
\bra{\phi _n}
{\rm U}_{12}^{\dagger} {\rm U}_{12} \ket{\Psi _a} \\
&=& -\bra{\phi _1} \otimes \bra{\phi _1} \otimes \bra{\phi _3 } \otimes \cdots \otimes
\bra{\phi _n} \ket{\Psi _a} \\
&=& 0,
\end{eqnarray*}
where ${\rm U}_{ij}$ is the swap operation between the $i$th and $j$th particle.
Extending this observation by induction we can easily show the following fact: We can always assume that
a state $\ket{\Phi _0} \stackrel{\rm def}{=} \ket{\phi _1} \otimes \cdots \otimes \ket{\phi _n}$ which
attains the maximum of $\max _{\ket{\Phi} \in {\rm Pro}(\Hi)}|\bra{\Phi }\ket{\Psi _a}|$ satisfies $\ket{\phi
_1} \perp \cdots \perp \ket{\phi _n}$. Then, under the condition of the orthogonality of the $\{ \ket{\phi _i}
\}_{i=1}^n$, we can calculate $|\bra{\Phi } \ket{\Psi _a}|$ as follows,
\begin{eqnarray*}
&\quad & |\bra{\phi _1} \otimes \cdots \otimes \bra{\phi _n} \ket{\Psi _a}| \\
&=&
|\bra{\phi _1} \otimes \cdots \otimes \bra{\phi _n} \ket{1} \wedge \cdots \wedge \ket{n}| \\
&=& |\frac{1}{\sqrt{n!}}\sum _{\sigma \in \mathfrak{G}_n} {\rm sign}(\sigma )\bra{\phi _1}\ket{\sigma
(1)}
\cdots \bra{\phi _n}\ket{\sigma (n)}    | \\
&=&     \frac{1}{\sqrt{n!}}|\det \{ \bra{\phi_i}\ket{j} \}_{ij} | \\
&=& \frac{1}{\sqrt{n!}},
\end{eqnarray*}
where $\{ \bra{\phi_i}\ket{j} \}_{ij}$ is a matrix with $\bra{\phi_i}\ket{j}$ as its $(i,j)$th element,
and we used the unitarity of $\{ \bra{\phi_i}\ket{j} \}_{ij}$ in the last equality. Therefore,
\begin{eqnarray*}
E_g (\ket{\Psi _a}) &=& - \log_2 \max _{\ket{\Phi} \in {\rm Pro}(\Hi)}|\bra{\Phi }\ket{\Psi _a}|^2 \\
                  &=& \log_2 n!.
\end{eqnarray*}
\hfill $\square$
\end{Proof}

Thus in the case of antisymmetric states we can derive the values of the other measures
from the value of geometric measure. That is, by Eq.(\ref{Eq geometric measure of anti-symmetric basis states})
and Eq.(\ref{equality for anti-symmetric}), we derive the following corollary.
\begin{Corollary}
In $\Hi = (\mathbb{C}^d)^{\otimes n}$, anti-symmetric basis states $\ket{\Psi _a} \stackrel{\rm def}{=}
\ket{l} \wedge \cdots \wedge \ket{l+n}$ satisfy
\begin{equation}\label{equality for anti-symmetric basis states}
\log _2 (1+R_g(\ket{\Psi _a})) = E_R(\ket{\Psi _a}) = E_g(\ket{\Psi _a}) = \log_2 n!,
\end{equation}
where $n \le d$, $1 \le l \le d-n$, and $\{ \ket{i} \}_{i=1}^d$ is an orthonormal basis on
$\mathbb{C}^d$.
\end{Corollary}

By Eq.(\ref{equality for anti-symmetric basis states}) and Eq.(\ref{robustness symmetric}), we can
compare the entanglement of anti-symmetric basis states $\ket{\Psi _a} = \ket{1}
\wedge \cdots \wedge \ket{n}$ with that of the symmetric basis states $\ket{\Psi _s} = \ket{S(n, (1, \cdots
,1))} = \sum _{\sigma \in \mathfrak{G}_n}\ket{\sigma(1)} \otimes \cdots \otimes \ket{\sigma(n)}$ on a
given Hilbert space $(\mathbb{C}^d)^{\otimes n}, \ (n \le d)$. Since $- \log_2 |\bra{1} \otimes \cdots \otimes
\bra{n} \ket{\Psi _s}|^2 = \log_2 n! = E_g(\ket{\Psi _a})$, we can easily see $E_g(\ket{\Psi _a}) \ge
E_g(\ket{\Psi _s})$, where equality holds if and only if $n =  2$. Moreover, when $n$ is
large enough, by means of Eq.(\ref{equality for anti-symmetric basis states}), Eq.(\ref{robustness symmetric})
and the Stirling formula, we derive
\begin{eqnarray}
\frac{E_g(\ket{\Psi _a})}{E_g(\ket{\Psi _s})} &\approx & \frac{n\log_2 n -n +1}{n +1} \nonumber \\
                                             &\approx &   \log_2 n.
\end{eqnarray}
Although the differences between anti-symmetric $\ket{\Psi _a}$ and symmetric $\ket{\Psi _s}$
basis states correspond only to phase factors ${\rm sign}(n)$, these two states have very different entanglement, and
actually an anti-symmetric basis state is more entangled than symmetric basis states. Furthermore,
since the symmetric basis states $\ket{\Psi _s} = \ket{S(n, (1, \cdots ,1))}$ have the largest values of
of entanglement among all symmetric basis states $\ket{S(\overrightarrow{k})}$ (under the
condition $n \le d$), the anti-symmetric basis states $\ket{\Psi _a}$ have larger values of the distance like
measures than all symmetric basis states in $(\mathbb{C}^d)^{\otimes n}$.




\section{Conclusion}\label{conclusion}
In this paper, we have discussed sufficient conditions under which the values of the distance like measures of
entanglement, ({\it i.e.} the robustness of entanglement, the relative entropy of entanglement, and the
geometric measure of entanglement), are equivalent by means of the representation theory of compact topological
groups (Theorem \ref{group representation} and Theorem \ref{group representation main}). As
applications of these theorems, we have seen that such distance like measures of entanglement are equivalent
for stabilizer states, projection states defined by the symmetric and anti-symmetric
subspaces (which include anti-symmetric basis states), and also for symmetric basis states. Moreover, by
calculating the value of the geometric measure of entanglement, we derived the values of all the
measures for anti-symmetric basis states and symmetric basis states. By comparing these values, we
conclude that anti-symmetric basis states are more entangled than any symmetric basis states on
$(\mathbb{C}^d)^{\otimes n}$ with $n \le d$. The results have applications as lower and upper bounds,
which can often be tight, on the optimal probability of discrimination by separable or LOCC
operations for certain classes of ensemble.

\section*{Acknowledgments}

We thank A. Miyake and F. Brand\~{a}o for very helpful discussions.
MH was supported by a MEXT Grant-in-Aid for Scientific Research on
Priority, Deepening and Expansion of Statistical Mechanical
Informatics (DEX-SMI), No. 18079014. SV was supported by the EPSRC
QIP-IRC, EU Integrated project QAP, the Royal Commission for the
Exhibition of 1851. MO and MM were supported by Special Coordination
Funds for Promoting Science and Technology. DM acknowledges support
from QICS.

\appendix

\section{Proof of Lemma 1}\label{appendixA}
\label{SEC: App symm}
\begin{Proof}
First, for an arbitrary $\ket{\Phi} \in \rm Pro (\Hi ^{\otimes n}) $, suppose that $\bra{\Phi}\ket{\Psi} = r e^{i
\theta}$, where $r \ge 0$ and $\theta$ is real. By choosing $\ket{\Phi'} \stackrel{\rm def}{=} e^{i
\theta} \ket{\Phi} \in \rm Pro (\Hi ^{ \otimes n})$, we can always find a state $\ket{\Phi'} $ such that
$\bra{\Phi'} \ket{\Psi}=|\bra{\Phi'}\ket{\Psi}| = |\bra{\Phi} \ket{\Psi}| =r$. Thus, when we consider
$\max _{\ket{\Phi}\in {\rm Pro(\Hi ^{\otimes n})} } |\langle \Phi|\Psi \rangle|$, we can always assume that
$\ket{\Phi}$ gives a non-negative real $\langle \Phi|\Psi \rangle$. In the following discussion, we always
assume $\ket{\Phi}$ satisfies this condition.

We prove this lemma in two steps; first for the case $n=2$ and later for
the case $n \geq 3$: \\

i) In the case $\ket{\Psi } \in \Hi ^{\otimes 2}$. \\

First, we note that the following proof is valid for non-normalized $\ket{\Psi}$. \\

We define $\ket{\Phi } \stackrel{\rm def}{=} \ket{a}\otimes \ket{b}$. A diagonalisation theorem known as
{\it Takagi's factorization} \cite{Horn and Johnson} states: ``{\it If $\Psi$ is a complex symmetric
matrix, then there exists a unitary ${\rm U}$ and a real nonnegative diagonal matrix $\Sigma =
\mathrm{diag}(r _1, \cdots , r_n )$ such that $\Psi = {\rm U} \Sigma {\rm U}^T$".} By means of this theorem, for
any $\ket{\Psi }$ in the symmetric subspace of $\Hi ^2$, we can calculate
\begin{eqnarray}
\bra{\Psi}\ket{a}\otimes\ket{b} &=& b^T \Psi a \nonumber\\
&=& ({\rm U}^T b)^T \Sigma {\rm U}^T a, \label{Takagi's factorization}
\end{eqnarray}
where in the first equality we used the natural correspondence between a bipartite Hilbert space and the
space of  matrices with respect to a fixed product basis: $\Psi$ is the $\dim \Hi \times \dim
\Hi$ matrix corresponding to $\ket{\Psi } \in \Hi \otimes \Hi$, and $a$ and $b$ are the column vectors
corresponding to $\ket{a}$ and $\ket{b}$, respectively. In Eq.(\ref{Takagi's factorization}), we also
note that ${\rm U}$ is a unitary matrix, and $\Sigma$ is a nonnegative diagonal matrix, both of which are
derived from Takagi's factorization. We can assume $r_1 \ge r_2 \ge \cdots \ge r_n$ for $\Sigma$. Then,
from Eq.(\ref{Takagi's factorization}), we can observe that the maximum of
$\bra{\Psi}\ket{a}\otimes\ket{b}$ is attained if and only if ${\rm U}^Tb = {\rm U}^Ta = e_1 \stackrel{\rm def}{=}( 1,
0, \cdots , 0 )^T$. Therefore, $\max _{\ket{a}\otimes \ket{b}}\bra{\Psi} \ket{a}\otimes \ket{b} = r_1$
and the maximum is attained if and only if $\ket{a} = \ket{b} = \ket{\overline{{\rm U}}e_1}$, where
$\ket{\overline{{\rm U}}e_1}$ is a state on $\Hi$ corresponding to the column vector $\overline{{\rm U}}e_1$,
($\overline{{\rm U}}$ is a complex conjugate of ${\rm U}$). Hence we have proven that Eq. (\ref{Eq geometric measure symmetric}) is
valid for bipartite states. \\

ii) In the case $\ket{\Psi} \in \Hi ^{\otimes n }, n \geq 3$. \\

Suppose $\ket{\Psi }$ is in the symmetric subspace of $\Hi ^{\otimes n}$, and assume that the state $\ket{a_1} \otimes \cdots \otimes
\ket{a_i} \otimes \cdots \otimes \ket{a_j } \otimes \cdots \otimes \ket{a_n}$ attains $\max
_{\ket{\Phi} \in {\rm Pro(\Hi ^{\otimes n})}} \bra{\Psi } \ket{\Phi} $ where $\ket{a_i } \neq
\ket{a_j}$. Then, ${\rm U}_{ij}\ket{\Psi } = \ket{\Psi} $ for all $ij$, where ${\rm U}_{ij}$ is the swap operation
between the ith and jth Hilbert spaces, i.e. the unitary defined as ${\rm U}_{ij}\ket{a_1} \otimes \cdots \otimes
\ket{a_i} \otimes \cdots \otimes \ket{a_j} \otimes \cdots \otimes \ket{a_n} = \ket{a_1} \otimes \cdots
\otimes \ket{a_j} \otimes \cdots \otimes \ket{a_i} \otimes \cdots \otimes \ket{a_n}$. Suppose $P_{ij}
\stackrel{\rm def}{=} \bra{a_1} \otimes \cdots \otimes \bra{a_{i-1}} \otimes I_{\Hi} \otimes
\bra{a_{i+1}} \otimes \cdots \otimes \bra{a_{j-1}} \otimes I_{\Hi} \otimes \bra{a_{j+1}} \otimes \cdots
\otimes \bra{a_n}$ is projection onto $\ket{a_1} \otimes \cdots \otimes \ket{a_{i-1}} \otimes \Hi
\otimes \ket{a_{i+1}} \otimes \cdots \otimes \ket{a_{j-1}} \otimes \Hi \otimes \ket{a_{j+1}} \otimes
\cdots \otimes \ket{a_n} \cong \Hi \otimes \Hi ,$ where $I_{\Hi }$ is the identity operator on $\Hi $.
Since ${\rm U}_{ij}P_{ij}\ket{\Psi} = P_{ij}{\rm U}_{ij}\ket{\Psi} = P_{ij}\ket{\Psi}$, $\ket{\Psi'} \stackrel{\rm
def}{=} P_{ij} \ket{\Psi}$ is a non-normalized symmetric bipartite state. By the definition of
$\ket{a_1} \otimes \cdots \otimes \ket{a_i} \otimes \cdots \otimes \ket{a_j } \otimes \cdots \otimes
\ket{a_n}$ and $\ket{\Psi'}$, we can obviously see that $\max _{\ket{a}\otimes \ket{b}}\bra{\Psi'}
\ket{a} \otimes \ket{b} = \max _{\ket{\Phi} \in {\rm Pro(\Hi ^{\otimes n)}}} \bra{\Psi}\ket{\Phi}$, and
$\ket{a_i} \otimes \ket{a_j}$ attains $\max _{\ket{a}\otimes \ket{b}}\bra{\Psi'} \ket{a} \otimes
\ket{b}$.

Then, from i), we can choose $\ket{a'_i}$ and $\ket{a'_j}$ such that $\ket{a'_i} = \ket{a'_j}$, and
$\ket{a'_i} \otimes \ket{a'_j}$ attains $\max _{\ket{a}\otimes \ket{b}}\bra{\Psi'} \ket{a} \otimes
\ket{b}$. That is, $\bra{\Psi'}\ket{a'_i}\otimes \ket{a'_j} =\bra{\Psi'} \ket{a_i} \otimes \ket{a_j} =
\max _{\ket{a}\otimes \ket{b}}\bra{\Psi'} \ket{a} \otimes \ket{b}$. Then, $\bra{\Psi} \ket{a_1} \otimes
\cdots \otimes \ket{a'_i} \otimes \cdots \otimes \ket{a'_j } \otimes \cdots \otimes \ket{a_n} =
\bra{\Psi} \ket{a_1} \otimes \cdots \otimes \ket{a_i} \otimes \cdots \otimes \ket{a_j } \otimes \cdots
\otimes \ket{a_n}$, and  $\ket{a_1} \otimes \cdots \otimes \ket{a'_i} \otimes \cdots \otimes \ket{a'_j
} \otimes \cdots \otimes \ket{a_n}$, which is symmetric for $(i,j)$, attains  $\max _{\ket{\Phi} \in
{\rm Pro(\otimes n)}} \bra{\Psi } \ket{\Phi}$.
 Therefore, by repeating the above symmetrization process for all $( i, j )$, we can
conclude that there always exists a $\ket{a_1} \otimes \cdots \otimes \ket{a_i} \otimes \cdots \otimes
\ket{a_j } \otimes \cdots \otimes \ket{a_n}$ which attains $ \max _{\ket{\Phi} \in {\rm Pro(\Hi
^{\otimes n})}} \bra{\Psi } \ket{\Phi} $, and also satisfies $\ket{a_i} = \ket{a_j}$ for all $(i,j)$.
 \hfill $\square$
\end{Proof}

\section{Elements of group representation theory} \label{appendix}

We first list the definitions and theorems that we use in the proof
of this paper.

\begin{Definition}{(Intertwining operator)}\label{definition of intertwining operator}
\\
Suppose $(\pi, \Hi )$ and $(\pi', \Hi')$ are both representations
of a group $G$. A linear operator $T$ from $\Hi$ onto $\Hi'$ is
called an intertwining operator if $T$ satisfies
\begin{equation}
\pi'(g) T = T \pi (g) \ (\forall g \in G).
\end{equation}
We write the set of all intertwining operators from $(\pi, \Hi )$
onto $(\pi', \Hi')$ as ${\rm Hom}_G (\Hi, \Hi')$.
\end{Definition}
${\rm Hom}_G(\Hi, \Hi')$ is a linear space.

\begin{Definition}{(Equivalence of group representations)}\label{definition of group equivalence}
\\
We say that two group representations $(\pi, \Hi)$ and $(\pi', \Hi')$ of
a group $G$ are equivalent, $(\pi, \Hi) \cong (\pi', \Hi')$, if
there exists a bijective linear map $A \in {\rm Hom}_G (\Hi,
\Hi')$.
\end{Definition}
In this case, $A$ gives an isomorphism between the group representations
$\pi(G)$ and $\pi'(G)$.

\begin{Definition}{(Multiplicity of irreducible representations)}
\label{multiplicity}
\\
Suppose a finite dimensional representation $(\pi, \Hi)$ of a group
$G$ is decomposed into a direct sum of irreducible representations
as $\Hi = \Hi _1 \oplus \cdots \oplus \Hi _k$. Then, for an
irreducible representation $(\tau, W)$ of $G$, it can be shown that $\dim
{\rm Hom}_G (W, \Hi) = \sharp \{i |(\tau, W) \cong (\pi | _{\Hi
_i}, \Hi _i) \}$. This dimension is called the {\it multiplicity} of
$\tau$ in $\pi$.
\end{Definition}

\begin{Lemma}{(Schur's lemma)} \label{Schur's lemma}
\\
Consider two given representations of a group $G$ on finite dimensional
complex Hilbert spaces, $(\pi, \Hi )$ and $(\pi', \Hi')$. If a
linear map $A: \Hi \rightarrow \Hi'$ satisfies,
\begin{equation}
A\pi (g) =\pi'(g)A, \ \forall g \in G,
\end{equation}
then, we have following.
\begin{enumerate}
\item If $(\pi, \Hi)$ and $(\pi', \Hi')$ are not equivalent, $A=0$
\item If $(\pi, \Hi) \cong (\pi', \Hi')$ and $T: \Hi \rightarrow
\Hi'$ gives an isomorphism, then, there exists $\lambda \in
\mathbb{C}$ such that $A = \lambda T$. In particular, in the case
$(\pi, \Hi) = (\pi', \Hi')$, $A = \lambda I$, where $I$ is the
identity on $\Hi$.
\end{enumerate}
\end{Lemma}
\begin{Proof}
See \cite{weyl, Iwanami}
\end{Proof}

\begin{Lemma} \label{invariant element in invariant subspace}
For a representation $(\pi, \Hi)$ of a group $G$, Suppose $\Hi$
can be decomposed as $\Hi = \oplus _{i=k}^K \Hi _k$, and each $\Hi
_k$ is invariant under the action of $G$. Then, $\Psi = \oplus _k
\Psi _k \in \Hi $ is an invariant element of $(\pi, \Hi)$ if and
only if $\Psi _k$ is an invariant element for all $k$.
\end{Lemma}
\begin{Proof}
The ``if part'' is trivial. \\
``only if part'': Suppose there exist $k_0$ such that $\pi (g)
\Psi _{k_0} \neq \Psi _{k_0}$. Then, from the uniqueness of the direct
sum decomposition, $\Psi = \oplus _{k=1}^K \Psi _k \neq \oplus
_{k=1}^K \pi (g) \Psi _k = \pi (g) \Psi$. This contradicts the
invariance of $\Psi $. \hfill $\square$
\end{Proof}
The following lemma, which concerns `averaging over' the Haar
measure of a compact topological group, is the key to deriving the
sufficient conditions under which the inequalities (\ref{Eqn: Ent ineq}) become equalities,
\begin{Lemma}\label{lemma iwanami} (See \cite{Iwanami}.)
Let $G$ be a compact topological group,  $(\pi, \Hi )$ a finite
dimensional unitary representation of the group $G$, and $dg$ a
normalized Haar measure on $G$. Then, the linear map on the
Hilbert-Schmidt space $\B _2 (\Hi) \cong \Hi \otimes \Hi
^{\dagger}$, (that is, the ``{\it super-operator }''),
\begin{equation}\label{definition of representation}
\rho \mapsto \int _G \pi (g) \rho \pi (g)^{\dagger } dg,
\end{equation}
is the projection (as a map on  $\B _2(\Hi )$) onto $\B_2(\Hi
)^{G}$, where $\B_2(\Hi )^{G}$ is the linear subspace of all
$G$-invariant elements on $\B _2 (\Hi )$;
\begin{equation}
\B_2(\Hi )^{G} \stackrel{\rm def}{=} \{ \rho \in \B _2(\Hi ) |
\forall g \in G, \ \pi (g) \rho \pi (g)^{\dagger} = \rho \}.
\end{equation}
\end{Lemma}
 All this lemma represents is that the integration (\ref{definition of
representation}) projects a state to the subspace of $G$-invariant
elements on $\B _2 (\Hi )$.


In cases where we know the irreducible decomposition of the group
representation $(\pi, \Hi)$, we can derive a concrete description
of the subspace of $G$-invariant elements $\B _2 (\Hi )^{G}$ as
follows. Since all compact topological groups are completely
reducible, $(\pi, \Hi)$ can be decomposed as
\begin{equation} \label{irreducible decomposition}
(\pi, \Hi) = (\bigoplus _{k=1}^K (I _{\mathcal A_k} \otimes \pi _k),
\bigoplus _{k=1}^K ({\mathcal A _k} \otimes {\mathcal B _k})),
\end{equation}
where $(\pi _k, {\mathcal B _k})$ is an irreducible representation
of the compact topological group $G$, and $(\pi _k, {\mathcal B
_k})$ and $(\pi _{k'}, {\mathcal B _{k'}})$ are inequivalent group representations for
all $k \neq k'$, i.e. there is no bijective
intertwining operator (see Definition \ref{definition of
intertwining operator} in this appendix) between the representation subspaces
corresponding to different $k$. In the above decomposition into
irreducible subspaces, we used a tensor product to write down
equivalent representations. By using this irreducible
representation, we can write down $\B_2(\Hi )^{G}$ explicitly as
follows. Note that the tensor product in Eq (\ref{irreducible
decomposition}) is not related to the tensor product of $\Hi =
\otimes _{i=1}^m \Hi _i$, which is the ``{\it cut}'' across which
we discuss the entanglement.
\begin{Lemma} \label{group representation 3}
For a given compact topological group $G$ and a unitary
representation on a finite dimensional complex Hilbert space $\Hi$,
$(\pi, \Hi)$,  we can write $\B_2(\Hi )^{G}$ as follows:
\begin{equation}
\B_2(\Hi )^{G} = \{ \bigoplus _{k=1}^K (N_k \otimes I_{{\mathcal
B_k}}) \in \B_2(\Hi )  | \forall k,  M_k \in \B _2({\mathcal A _k})
\},
\end{equation}
where $\mathcal{A}_k$ and $\mathcal{B}_k$ are defined by the
irreducible decomposition $(\pi, \Hi ) = (\bigoplus _{k=1}^K (I
_{\mathcal{A}_k} \otimes \pi _k), \bigoplus _{k=1}^K (\mathcal{A}_k
\otimes \mathcal{B}_k))$.
\end{Lemma}
\begin{Proof}
As with Theorem \ref{group representation}, we consider the
unitary representation of $G$ on $\B _2 (\Hi) \cong \Hi \otimes
\Hi ^{\dagger}$ via the map $\rho \mapsto \pi(g) g \pi(g)^{\dagger}$. We
denote this representation by $(\pi \otimes \pi ^{\dagger }, \Hi
\otimes \Hi ^{\dagger} )$. Since a compact topological group is
completely reducible, this representation can be decomposed as
$(\pi, \Hi)=(\bigoplus _{k=1}^K (I _{\mathcal{A}_k} \otimes \pi _k),
\bigoplus _{k=1}^K (\mathcal{A}_k \otimes \mathcal{B}_k))$, where
$(\pi _k, \mathcal{A}_k \otimes \mathcal{B}_k)$ is irreducible for
all $k$, and $(\pi _k, \mathcal{A}_k \otimes \mathcal{B}_k)$ and
$(\pi _{k'}, \mathcal{A}_{k'} \otimes \mathcal{B}_{k'})$ are not
equivalent for all $k \neq k'$. Then the representation on the
Hilbert Schmidt space $\Hi \otimes \Hi ^{\dagger }$ also
decomposes as
\begin{widetext}
\begin{eqnarray*}
(\pi  \otimes \pi ^{\dagger }, \Hi \otimes \Hi ^{\dagger}) &= &
(\bigoplus _{k,l} (I _{\mathcal{A}_k} \otimes \pi _k \otimes I
_{\mathcal{A}_l}^{\dagger} \otimes \pi _l^{\dagger}),
 \bigoplus _{k,l} (\mathcal{A}_k \otimes \mathcal{B}_k \otimes \mathcal{A}_l^{\dagger} \otimes \mathcal{B}_l^{\dagger})) \\
&\cong & (\bigoplus _{k,l} (I _{\mathcal{A}_k} \otimes I
_{\mathcal{A}_l}^{\dagger} \otimes \pi _k  \otimes \pi
_l^{\dagger}),
 \bigoplus _{k,l} (\mathcal{A}_k \otimes \mathcal{A}_l^{\dagger} \otimes \mathcal{B}_k  \otimes \mathcal{B}_l^{\dagger})),
\end{eqnarray*}
\end{widetext}
where in the second line we have reordered the tensor spaces for
convenience in later discussions.

That is, each $ (\mathcal{A}_k \otimes \mathcal{A}_l^{\dagger}
\otimes \mathcal{B}_k  \otimes \mathcal{B}_l^{\dagger})$, (that is,
the Hilbert Schmidt space of operators between $\mathcal{A}_k  \otimes \mathcal{B}_k
$ and $\mathcal{A}_l  \otimes \mathcal{B}_l $), is a invariant
subspace of $\pi \otimes \pi ^{\dagger}$ for any $k$ and $l$. Then,
from Lemma \ref{invariant element in invariant subspace} in appendix
\ref{appendix}, in order to derive the description of an invariant
element of $(\pi \otimes \pi ^{\dagger }, \Hi \otimes \Hi ^{\dagger}
)$, we only need to consider the invariant element of $ (I
_{\mathcal{A}_k} \otimes I _{\mathcal{A}_l}^{\dagger} \otimes \pi _k
\otimes \pi _l^{\dagger}, (\mathcal{A}_k \otimes
\mathcal{A}_l^{\dagger} \otimes \mathcal{B}_k  \otimes
\mathcal{B}_l^{\dagger}))$, and the invariant element of the whole
space is only the direct product of such invariant elements of
subspace. That is, $M \in \Hi \otimes \Hi ^{\dagger}$ is an
invariant element, if and only if $M= \oplus _{k,l} M_{kl}$ and all
$M_{kl} \in \mathcal{A}_k \otimes \mathcal{A}_l^{\dagger} \otimes
\mathcal{B}_k   \otimes \mathcal{B}_l^{\dagger}$ are invariant
elements.

Suppose $M_{kl} \in \mathcal{A}_k \otimes \mathcal{A}_l^{\dagger}
\otimes \mathcal{B}_k  \otimes \mathcal{B}_l^{\dagger}$ is an
invariant element for all $k,l$. Then, from $\pi (g) M_{kl} \pi
(g)^{\dagger} = I _{\mathcal{A}_k} \otimes \pi _k (g) M_{kl} I
_{\mathcal{A}_l} \otimes \pi _l^{\dagger}(g)$, we derive
\begin{equation} \label{intertwining operator}
\pi _k (g) \bra{\alpha _p^k}M_{kl}\ket{\alpha _q^l} =  \bra{\alpha
_p^k}M_{kl}\ket{\alpha _q^l} \pi _l(g)
\end{equation}
for all $kl$ and $g \in G$, where $\{ \ket{\alpha _p^k} \}_{p
=1}^{\dim \mathcal{A}_k}$ and $\{ \ket{\alpha _q^l} \}_{q =1}^{\dim
\mathcal{A}_l}$ are orthonormal basis of $\mathcal{A}_k$ and
$\mathcal{A}_l$, respectively. Here, we should note $\bra{\alpha
_p^k}M_{kl}\ket{\alpha _q^l} \in  \mathcal{B} _k \otimes
\mathcal{B}_l^{\dagger}$ for all $p,q$. Next, we use Schur's lemma
(Lemma \ref{Schur's lemma} in this appendix) for the representation
$(\pi _k, \mathcal{B} _k)$ and $(\pi _l, \mathcal{B}_l)$. Then,
since $(\pi _k, \mathcal{B} _k)$ and $(\pi _l, \mathcal{B} _l)$ are
not equivalent for all $k \neq l$, by means of Eq.(\ref{intertwining
operator}) and Schur's lemma, we derive
\begin{eqnarray}
\bra{\alpha _p^k}M_{kl}\ket{\alpha _q^l} =0 \qquad (\forall k \neq l, {\rm and }\ \forall p,q ) \label{application of shur 1}\\
\bra{\alpha _p^k}M_{kl}\ket{\alpha _q^k} = C_{pq}^k I_{\mathcal{B}
_k} \qquad (\forall k, {\rm and }\ \forall p,q ), \label{application
of shur 2}
\end{eqnarray}
where $C_{pq} \in \mathbb{C}$ is a complex number coefficient. By
using matrix element $m_{pqrs}$, $M_{kl}$ can be written as $M_{kl}
= \sum _{pqrs} m_{pqrs}^{kl} \ket{\alpha _p^k}\bra{\alpha _q^l}
\otimes \ket{\beta _r^k}\bra{\beta _s^l}$, where $\{ \ket{\beta
_r^k} \}_{r =1}^{\dim \mathcal{B}_k}$ and $\{ \ket{\beta _r^l} \}_{r
=1}^{\dim \mathcal{B}_l}$ are orthonormal bases of $\mathcal{B}_k$
and $\mathcal{B}_l$, respectively. Then, Eq. (\ref{application of
shur 1}) and Eq. (\ref{application of shur 2}) can be written down
as
\begin{eqnarray}
m_{pqrs}^{kl} =0 \qquad (\forall k \neq l, {\rm and} \ \forall  p,q,r,s ) \label{application of shur 3}\\
m_{pqrs}^{kk} = C_{pq} \delta _{rs}  \qquad (\forall k, {\rm and }\
\forall p,q ), \label{application of shur 4}.
\end{eqnarray}
From Eq. (\ref{application of shur 3}), we derive $M_{kl} = 0$ for
all $k \neq l$. From  Eq. (\ref{application of shur 4}), we
derive
\begin{eqnarray*}
M_{kk} &=&  (\sum _{pq} C_{pq}^k \ket{ \alpha _p^k} \bra{\alpha _q^k}) \otimes (\sum _r \ket{\beta _r^k}\bra{\beta _r^k}) \\
       &=&  N_k \otimes I_{\mathcal{B}_k},
\end{eqnarray*}
where $N_k \stackrel{\rm def}{=} \sum _{pq} C_{pq}^k \ket{ \alpha
_p^k} \bra{\alpha _q^k} \in \mathcal{A}_k \otimes
\mathcal{A}_k^{\dagger}$.

Therefore, finally by means of Lemma \ref{invariant element in
invariant subspace}, we can conclude that $M \in \Hi \otimes \Hi
^{\dagger}$ is an invariant element of $(\pi \otimes \pi ^{\dagger},
\Hi \otimes \Hi ^{\dagger})$ if $M$ can be written down as $M =
\bigoplus _{k=1}^K (N_k \otimes I_{\mathcal{B}_k})$ by using $N_k
\in \mathcal{A}_k \otimes \mathcal{A}_k^{\dagger}$. Conversely,
suppose $M \in \Hi \otimes \Hi ^{\dagger}$ can be written as $M
= \bigoplus _{k=1}^K (N_k \otimes I_{\mathcal{B}_k})$. Then, since
all $N_k \otimes I_{\mathcal{B}_k}$ are clearly invariant elements
of $(\pi \otimes \pi ^{\dagger}, \Hi \otimes \Hi ^{\dagger})$,  by
Lemma \ref{invariant element in invariant subspace} in this appendix,
$M$ is also an invariant element. \hfill $\square$
\end{Proof}

By means of the previous lemma, we can derive an expression for
the state that results from averaging over a compact topological group as follows.
\begin{Lemma} \label{group representation extra 1}
For a given compact topological group $G$ and a corresponding unitary
representation on a finite dimensional complex Hilbert space $\Hi$,
$(\pi, \Hi)$, a projection (super-) operator $\mathcal{P} _{\B
_2(\Hi)^G}$ onto the Hilbert-Schmidt subspace of $G$-invariant
elements $\B _2(\Hi)^G$ maps a Hilbert Schmidt class operator $\rho
\in \B _2(\Hi )$ as follows,
\begin{equation}
\mathcal{P}_{\B _2(\Hi)^G} (\rho )= \sum _{k=1}^K \frac{1}{\dim
\mathcal{B}_k}\Tr _{\mathcal{B}_k}(P_{\mathcal{A}_k \otimes
\mathcal{B}_k} \rho P_{\mathcal{A}_k \otimes \mathcal{B}_k}) \otimes
I_{\mathcal{B}_k}
\end{equation}
where $\mathcal{A}_k$ and $\mathcal{B}_k$ is defined by the
irreducible decomposition $(\pi, \Hi ) = (\bigoplus _{k=1}^K (I
_{\mathcal{A}_k} \otimes \pi _k), \bigoplus _{k=1}^K (\mathcal{A}_k
\otimes \mathcal{B}_k))$, and $P_{\mathcal{A}_k \otimes
\mathcal{B}_k}$ is a projection onto $\mathcal{A}_k \otimes
\mathcal{B}_k$.
\end{Lemma}
\begin{Proof}
From Lemma \ref{group representation 3}, $\B _2(\Hi )^G$ can be
written down as $B _2(\Hi )^G = \bigoplus _{k=1}^K (\B
_2(\mathcal{A}_k ) \otimes \{ \alpha I_{\mathcal{B}_k} \}_{\alpha
\in \mathbb{C}})$, where $I _{\mathcal{B}_k}$ is the identity
operator on subspace $\mathcal{B}_k$.

Suppose $\mathcal{P} _{I _{\mathcal{B}_k}}$ is a projection onto a
one-dimensional Hilbert-Schmidt subspace $\{ \alpha
I_{\mathcal{B}_k} \}_{\alpha \in \mathbb{C}}$. Then, for $\rho \in
\B _2 (\mathcal{B}_k)$, $\mathcal{P} _{I _{\mathcal{B}_k}}(\rho )=
\frac{\Tr (\rho)}{\dim \mathcal{B}_k} \cdot I_{\mathcal{B}_k}$ Thus,
\begin{eqnarray*}
&\quad & \mathcal{P}_{\B _2(\Hi)^G} (\rho ) \\
&=& \sum _{k=1}^K \mathcal{P}_{\B
_2(\mathcal{A}_k) \otimes \{ \alpha I_{\mathcal{B}_k} \}_{\alpha \in
\mathbb{C}}}
\circ \mathcal{P}_{\B _2 (\mathcal{A}_k \otimes \mathcal{B}_k)} ( \rho ) \\
&=& \sum _{k=1}^K (\mathcal{P}_{\B _2(\mathcal{A}_k)} \otimes
\mathcal{P}_{\{ \alpha I_{\mathcal{B}_k} \}_{\alpha \in \mathbb{C}}})\cdot (P_{\mathcal{A}_k \otimes \mathcal{B}_k} \rho P_{\mathcal{A}_k \otimes \mathcal{B}_k}) \\
&=& \sum _{k=1}^K \frac{1}{\dim \mathcal{B}_k} \Tr _{\mathcal{B}}
(P_{\mathcal{A}_k \otimes \mathcal{B}_k} \rho P_{\mathcal{A}_k
\otimes \mathcal{B}_k})\otimes I_{\mathcal{B}_k}
\end{eqnarray*}
\hfill $\square$
\end{Proof}

In the next appendix we will attempt to apply the above definitions and
lemmas, with the intention of deriving a fairly general sufficient conditions under which equality of the measures may be proven.
However, it turns out that these sufficient conditions are equivalent to a more obvious sufficient condition -
that the state under consideration is the invariant state of an irreducible subspace of multiplicity one in the local unitary representation.
Nevertheless, we present the full arguments
below, in order that the origin of this condition be clear.



\section{Sufficient conditions under which distance like measures of entanglement coincide}\label{more general group discussion}

In this appendix, we present the sufficient conditions under which
the measures of entanglement that we consider coincide by means of the
group theoretical tools reviewed above.

As stated previously, these conditions
collapse to the more elementary condition that multiplicity of the irrep upon which the invariant state
resides is 1. That this is a sufficient condition may be seen more directly by an intuitive
argument - the off-diagonal elements representing coherences between inequivalent irreps must
 vanish as a consequence of Schur's lemma, hence if the invariant state is proportional to
 the projector onto an irrep of multiplicity 1, the closest product state under the geometric measure
 automatically averages under twirling to give a state of the form required for the robustness measure. Hence a similar discussion
 holds in such examples as with the stabilizer states. However, we present the full sequence of lemmas as
 the conditions that we develop at first seem to be more general, and so it is of interest to understand
 why this is not the case.

First, by means of Lemma \ref{lemma iwanami}, we derive the following
sufficient condition under which the distance like measures have
equal value.
\begin{Lemma}\label{group representation}
A projection state $\frac{P}{\Tr P }$ on $\Hi = \otimes _{i=1}^m \Hi
_i$ satisfies,
\begin{equation} \label{equality of distance like measure}
\log _2 (1+R_g(\frac{P}{\Tr P})) = E_R(\frac{P}{\Tr P}) =
G(\frac{P}{\Tr P})-\log_2 \Tr P,
\end{equation}
if there exists a compact topological group $G$ and a finite
dimensional unitary representation $(\pi ,  \Hi )$ such that $P$ is
an invariant element of the representation of $G$ as defined by
Eq.(\ref{definition of representation}), $\pi(g )$ is a local
unitary transformation for all $g \in G$, and the following
inequality is satisfied,
\begin{equation}\label{inequality of group representation}
\int _G \pi (g)\ket{\Phi _0}\bra{\Phi _0}\pi (g)^{\dagger} dg \ge
 \frac{\bra{\Phi _0}P\ket{\Phi _0}}{\Tr P} P,
\end{equation}
where $\ket{\Phi _0}$ attains $\max _{\ket{\Phi} \in Pro(\Hi )}
\bra{\Phi}\frac{P}{\Tr P}\ket{\Phi}$.
\end{Lemma}
\begin{Proof}
Suppose $\frac{P}{\Tr P}$ is an invariant element of the
representation of $G$,  $\pi(g)$ is a local unitary for all $g \in
G$, and the inequality (\ref{inequality of group representation})
holds. Then, from Lemma \ref{lemma iwanami},
\begin{eqnarray}
 &\quad& \int _G \pi (g) \ket{\Phi _0} \bra{\Phi _0} \pi (g)^{\dagger} dg
\nonumber \\
&=&
\mathcal{P}_{\B _2( \Hi )^G }(\ket{\Phi _0}\bra{\Phi _0}) \nonumber \\
&=& \mathcal{P}_{P}(\ket{\Phi _0}\bra{\Phi _0}) +
\mathcal{P}_{P{\perp}} (\ket{\Phi
_0}\bra{\Phi _0}) \nonumber \\
&=&  \bra{\Phi_0} \frac{P}{\sqrt{\Tr P}}  \ket{\Phi_0}
\frac{P}{\sqrt{\Tr P}} + \mathcal{P}_{P^{\perp}} (\ket{\Phi
_0}\bra{\Phi
_0})  \in  {\rm Sep}( \Hi ) \nonumber \\
&=&  (\Tr P)\bra{\Phi_0} \frac{P}{\Tr P}  \ket{\Phi_0} \frac{P}{\Tr
P}  \nonumber \\
&\quad & + \mathcal{P}_{P^{\perp}} (\ket{\Phi _0}\bra{\Phi _0})   \in
{\rm Sep}( \Hi ), \label{inequality main lemma}
\end{eqnarray}
where $\mathcal{P}_{\B _2(\Hi )^G }$, $\mathcal{P}_{P}$, and
$\mathcal{P}_{P^{\perp}}$ are the projections of Hilbert-Schmidt
space (super-operator) $\B _2(\Hi )$ onto $\B _2( \Hi )^G$, $P$, and
the orthogonal complement of $P$ as a subspace of $\B _2( \Hi )^G $,
respectively.

By definition, $\mathcal{P}_{P^{\perp}}(\ket{\Phi _0}\bra{\Phi _0})$
satisfies $ \Tr P \ \mathcal{P}_{P^{\perp}}(\ket{\Phi _0}\bra{\Phi
_0}) = 0 $, and from (\ref{inequality of group representation}),
$\mathcal{P}_{P^{\perp}}(\ket{\Phi _0}\bra{\Phi _0}) \ge 0$. Hence,
from Eq.(\ref{Eqn: DEF robustness}) and Eq.(\ref{inequality main
lemma}), by corresponding  $\omega$ and $\frac{1}{1+t} $ to $\int _G \pi
(g) \ket{\Phi _0} \bra{\Phi _0} \pi (g)^{\dagger} dg$ and
$\bra{\Phi_0} P \ket{\Phi_0}$, we see that
$\frac{1}{\bra{\Phi_0} P \ket{\Phi_0}}-1$ satisfies the all
condition of $t$ in the definition (\ref{Eqn: DEF robustness}) of $R_g(\rho)$. Thus, we derive $\frac{1}{\bra{\Phi
_0}P\ket{\Phi _0}} \ge 1+R_g(\frac{P}{\Tr P})$. Moreover,
by the definition of $\ket{\Phi_0}$, $\frac{1}{\bra{\Phi
_0}P\ket{\Phi _0}} = 2^{G(\frac{P}{\Tr P})-\log_2 \Tr P}$. That is,
\begin{eqnarray}
1+R_g(\frac{P}{\Tr P}) \le \frac{1}{\bra{\Phi _0}P\ket{\Phi _0}} &=& 2^{G(\frac{P}{\Tr P})-\log_2 \Tr P} \nonumber \\
&\le & 1+R_g(\frac{P}{\Tr P}), \label{inequality for proof of theorem 1}
\end{eqnarray}
where we use the inequalities (\ref{Eqn: Ent ineq}) in the second
inequality. Therefore, $G(\frac{P}{\Tr P})-\log_2 \Tr P = E_r(\frac{P}{\Tr P}) = \log _2
(1+R_g(\frac{P}{\Tr P}))$. \hfill $\square$
\end{Proof}
In the above Theorem, the inequality (\ref{inequality of group
representation}) corresponds to the condition that $\int _G \pi
(g)\ket{\Phi _0} \bra{\Phi _0} \pi(g)^{\dagger} dg$ should be in the
form $\int _G \pi (g)\ket{\Phi _0} \bra{\Phi _0} \pi(g)^{\dagger} dg
= \lambda (\Tr P) \frac{P}{\Tr P} + \{1-\lambda(\Tr P)\} \Delta$
with a positive $\lambda {\Tr P} \le 1$ and a state $\Delta$, which we used
in the definition of $R_g(\frac{P}{\Tr P})$ (\ref{Eqn: DEF robustness}).
This condition is necessary for this upper bound (the first
inequality in Eq.(\ref{inequality for proof of theorem 1})) to be
valid.

From the proof of the above lemma,
we can easily see that $\int _G \pi (g)
\ket{\Phi _0}\bra{\Phi _0} \pi (g)^{\dagger} dg $ is a closest
separable state of a projection state $\frac{P}{\Tr P}$ in terms of the robustness of entanglement
in the case where the projection state satisfies all of the assumption in the above lemma.
Moreover, we can also show that this state is a closest separable state
in terms of the relative entropy of entanglement.
Hence, roughly speaking, if a given multipartite projection state has enough group
symmetry, we can derive a closest separable state from a closest
product state by just averaging it over a group action.
We can check the optimality for the relative entropy of entanglement
by the following argument, which was also used in \cite{Hayashi05}.
\begin{eqnarray*}
&\quad & E_R(\frac{P}{\Tr P}) \\
& = & \min _{\omega \in {\rm Sep} } D(\rho \| \omega) \\
& \leq &  D(\rho \| \int _G \pi (g) \ket{\Phi _0} \bra{\Phi _0} \pi (g)^{\dagger} dg) \\
& = & -\log_2 \Tr P  \\
&\quad & - \Tr \left ( \frac{P}{\Tr P} \log_2 \left (
\frac{\bra{\Phi _0} P \ket{\Phi _0} }{\Tr P}P +
\mathcal{P}_{P^{\perp}}(\ket{\Phi _0}\bra{\Phi _0}) \right) \right ) \\
& \le & -\log_2 \Tr P - \Tr \left ( \frac{P}{\Tr P} \log_2 \left ( \frac{\bra{\Phi _0} P \ket{\Phi _0} }{\Tr
 P}P  \right ) \right ) \\
& = & - \log_2 \ket{\Phi _0} P \bra{\Phi _0} \\
&=& G(\frac{P}{\Tr P}) -\log_2 \Tr P,
\end{eqnarray*}
where we used the operator-monotonicity of the logarithmic function in the second inequality.
Since $E_R(\frac{P}{\Tr P}) = G(\frac{P}{\Tr P}) -\log_2 \Tr P $, all
of the above inequalities should be equalities. Therefore, from the
second equality, we derive $\min _{\omega \in {\rm Sep} } D(\rho \|
\omega)
 =  D(\rho \| \int _G \pi (g) \ket{\Phi _0} \bra{\Phi _0} \pi (g)^{\dagger} dg)$;
that is, $\int _G \pi (g) \ket{\Phi _0} \bra{\Phi _0} \pi
(g)^{\dagger} dg$ is a closest separable state in terms of the
relative entropy.

By means of Lemma \ref{group representation extra 1}, we can
rewrite Lemma \ref{group representation} as follows.
\begin{Theorem}\label{group representation extra 2}
A projection state $\frac{P}{\Tr P }$ on $\Hi = \otimes _{i=1}^m \Hi
_i$ satisfies,
\begin{equation} \label{equality of distance like measure extra 2}
\log _2 (1+R_g(\frac{P}{\Tr P})) = E_R(\frac{P}{\Tr P}) =
G(\frac{P}{\Tr P})-\log_2 \Tr P,
\end{equation}
if there exist a compact topological group $G$ and its finite
dimensional unitary representation $(\pi ,  \Hi )$ such that $P$ is
a $G$-invariant element, $\pi(g )$ is a local unitary transformation
for all $g \in G$, and the following inequality is satisfied for all
$k$ such that ${\rm Ran}(P) \cap (\mathcal{A}_k \otimes
\mathcal{B}_k) \neq \{ 0 \}$,
\begin{eqnarray}\label{inequality of group representation extra 2}
&& \Tr _{\mathcal{B}_k}(P_{\mathcal{A}_k \otimes \mathcal{B}_k}
\ket{\Phi _0}\bra{\Phi _0}P_{\mathcal{A}_k \otimes \mathcal{B}_k})
\nonumber \\
&&-
\frac{\bra{\Phi _0} P \ket{\Phi _0}}{\Tr P} \Tr
_{\mathcal{B}_k}(P_{\mathcal{A}_k \otimes \mathcal{B}_k} P
P_{\mathcal{A}_k \otimes \mathcal{B}_k}) \ge 0,
\end{eqnarray}
where  $\mathcal{A}_k$ and $\mathcal{B}_k$ are define by the
irreducible decomposition $(\pi, \Hi ) = (\bigoplus _{k=1}^K (I
_{\mathcal{A}_k} \otimes \pi _k), \bigoplus _{k=1}^K (\mathcal{A}_k
\otimes \mathcal{B}_k))$,  $P_{\mathcal{A}_k \otimes \mathcal{B}_k}$
is a projection onto $\mathcal{A}_k \otimes \mathcal{B}_k$, and
$\ket{\Phi _0}$ attains $\max _{\ket{\Phi} \in Pro(\Hi )}
\bra{\Phi}\frac{P}{\Tr P}\ket{\Phi}$.
\end{Theorem}
\begin{Proof}
Suppose all assumptions in this theorem are valid. Then,
\begin{eqnarray*}
&\quad & \int _G \pi (g)\ket{\Phi _0}\bra{\Phi _0}\pi (g)^{\dagger}
dg -
\frac{\bra{\Phi _0} P \ket{\Phi _0}}{\Tr P}   P \\
&=& \int _G \pi (g) \left ( \ket{\Phi _0}\bra{\Phi _0} - \frac{\bra{\Phi _0} P \ket{\Phi _0}}{\Tr P} P \right ) \pi (g)^{\dagger} dg \\
&=& \sum _{k=1}^K \Tr _{\mathcal{B}_k} \Big(P_{\mathcal{A}_k
\otimes \mathcal{B}_k} \big( \ket{\Phi _0}\bra{\Phi _0}
 \\
&\quad & \qquad - \frac{\bra{\Phi _0} P \ket{\Phi _0}}{\Tr P}  P \big) P_{\mathcal{A}_k \otimes \mathcal{B}_k} \Big) \otimes I_{\mathcal{B}_k} \\
&=& \sum _{k=1}^K \Big(\Tr _{\mathcal{B}_k}(P_{\mathcal{A}_k
\otimes \mathcal{B}_k} \ket{\Phi _0}\bra{\Phi _0}P_{\mathcal{A}_k
\otimes \mathcal{B}_k})  \\
&\quad &\qquad - \frac{\bra{\Phi _0} P \ket{\Phi _0}}{\Tr
P} \Tr _{\mathcal{B}_k} \left (P_{\mathcal{A}_k \otimes
\mathcal{B}_k} P P_{\mathcal{A}_k \otimes \mathcal{B}_k} \right )
\Big) \otimes I_{\mathcal{B}_k},
\end{eqnarray*}
where we used Lemma \ref{group representation extra 1} in the second
equality. Thus, $\int _G \pi (g)\ket{\Phi _0}\bra{\Phi _0}\pi
(g)^{\dagger} dg - \frac{\bra{\Phi _0} P \ket{\Phi _0}}{\Tr P}   P
\ge 0$, if and only if $\Tr _{\mathcal{B}_k}(P_{\mathcal{A}_k
\otimes \mathcal{B}_k} \ket{\Phi _0}\bra{\Phi _0}P_{\mathcal{A}_k
\otimes \mathcal{B}_k}) - \frac{\bra{\Phi _0} P \ket{\Phi _0}}{\Tr
P} \Tr _{\mathcal{B}_k} \left (P_{\mathcal{A}_k \otimes
\mathcal{B}_k} P P_{\mathcal{A}_k \otimes \mathcal{B}_k} \right )
\ge 0$ for all $k $ such that ${\rm Ran}(P) \cap (\mathcal{A}_k
\otimes \mathcal{B} _k) \neq \{ 0 \}$. Therefore, from Lemma
\ref{group representation}, we can derive this theorem. \hfill
$\square$
\end{Proof}

For a pure state, the sufficient condition in the above theorem can be
simplified to the following.
\begin{Theorem}\label{group representation for pure}
a state $\ket{\Psi} \in \Hi = \otimes _{i =1}^m \Hi _i$ satisfies,
\begin{equation} \label{equality of distance like measure for pure}
\log _2 (1+R_g(\ket{\Psi })) = E_R(\ket{\Psi}) = E_g(\ket{\Psi}),
\end{equation}
if there exists a compact topological group $G$ and its finite
dimensional unitary representation $(\pi , \Hi )$ such that
$\ket{\Psi } \bra{\Psi}$ is a $G$-invariant element, $\pi(g )$ is a
local unitary transformation for all $g \in G$, and the following
inequality is satisfied,
\begin{equation}\label{inequality of group representation for pure}
P_{\mathcal{A}_{k_0} \otimes \mathcal{B}_{k_0}}\ket{\Phi
_0}\bra{\Phi _0}P_{\mathcal{A}_{k_0} \otimes
\mathcal{B}_{k_0}}-|\langle{\Psi}|{\Phi _0}\rangle |^2 \ket{\Psi}\bra{\Psi}
\ge 0,
\end{equation}
where  $\mathcal{A}_k$ and $\mathcal{B}_k$ are defined by the
irreducible decomposition $(\pi, \Hi ) = (\bigoplus _{k=1}^K (I
_{\mathcal{A}_k} \otimes \pi _k), \bigoplus _{k=1}^K (\mathcal{A}_k
\otimes \mathcal{B}_k))$,  $P_{\mathcal{A}_k \otimes \mathcal{B}_k}$
is a projection onto $\mathcal{A}_k \otimes \mathcal{B}_k$, $k_0$
satisfies $ \mathcal{A}_{k_0} \otimes \mathcal{B}_{k_0} \ni
\ket{\Psi} $, and $\ket{\Phi _0}$ attains $\max _{\ket{\Phi} \in
Pro(\Hi )} \bra{\Phi}\frac{P}{\Tr P}\ket{\Phi}$.
\end{Theorem}

In the above theorem \ref{group representation}, in order to check
whether a projection state $\frac{P}{\Tr P}$ satisfies
Eq.(\ref{inequality of group representation}), or not, we need to
know the closest product state $\ket{\Phi _0}$, that is, the state
which attain $\max _{\Phi \in \rm Pro (\Hi)} \bra{\Phi} \frac{P}{\Tr
P} \ket{\Phi}$. However, if $\frac{P}{\Tr P}$ and a group
representation $(\pi, \Hi)$ of a topological group $G$ satisfy an
additional condition, we can derive Eq.(\ref{equality of distance
like measure}) without needing to know the closest product state
$\ket{\Phi _0}$. We can write down this fact as following lemma.
\begin{Lemma}\label{group representation 2}
a projection state $\frac{P}{\Tr P}$ on $\Hi = \otimes _{i=1}^m \Hi
_i$ satisfies,
\begin{equation} \label{equality of distance like measure 2}
\log _2 (1+R_g(\frac{P}{\Tr P})) = E_R(\frac{P}{\Tr P}) =
G(\frac{P}{\Tr P})-\log_2 \Tr P,
\end{equation}
if there exists a compact topological group $G$ and its finite
dimensional unitary representation $(\pi , \Hi)$ such that $P \in
\B_2(\Hi )^{G}$ , $\pi(g )$ is a local unitary transformation for
all $g \in G$, and $\sigma \ket{\xi} = 0$ for all $\ket{\xi} \in
{\rm Ran}(P) $ and $\sigma \in \B ^G _{P^{\perp}}$, where ${\rm
Ran}(P)$ is the range (the image of the domain) of the projection
$P$, and $\B ^G _{P{\perp}}$ is defined as an orthogonal complement
of $P$ in the Hilbert-Schmidt subspace $\B _2(\Hi)^G $,
\begin{equation}
\B ^G _{P^{\perp}} \stackrel{\rm }{=} \{ \sigma \in \B _2(\Hi )^G
|\Tr P \sigma =0 \}.
\end{equation}
\end{Lemma}
\begin{Proof}
We will see that, if $\sigma \ket{\xi} = 0$ for all $\ket{\xi} \in
{\rm Ran}(P)$ and $\sigma \in \B ^G _{P^{\perp}}$, the inequality
(\ref{inequality of group representation}) is satisfied. Suppose all
conditions of this lemma are satisfied. Then, by using the same
discussion as that of Theorem \ref{group representation}, $\int _G
\pi (g) \ket{\Phi _0}\bra{\Phi _0} \pi (g)^{\dagger}dg$ can be
written down as,
\begin{equation} \label{equation 1005}
\int _G \pi (g) \ket{\Phi _0}\bra{\Phi _0} \pi (g)^{\dagger} dg=
 \frac{\bra{\Phi _0}P \ket{\Phi _0}}{\Tr P}P
+ \mathcal{P} _{P^{\perp}}(\ket{\Phi _0}\bra{\Phi _0}).
\end{equation}
Since $\Tr P \mathcal{P} _{\ket{\xi}\bra{\xi}^{\perp}}(\ket{\Phi
_0}\bra{\Phi _0}) =0 $, $\mathcal{P}
_{\ket{\xi}\bra{\xi}^{\perp}}(\ket{\Phi _0}\bra{\Phi _0}) \in
\B^G_{\Psi ^{\perp}}$. Then,  by the assumption of this lemma,
$\mathcal{P} _{\ket{\xi}\bra{\xi}^{\perp}}(\ket{\Phi _0}\bra{\Phi
_0})\ket{\xi} =0$ for all $\ket{\xi} \in {\rm Ran}(P)$. Therefore,
for all $\ket{\xi } \in {\rm Ran } (P)$,
\begin{equation}
\int _G \pi (g) \ket{\Phi _0}\bra{\Phi _0} \pi (g)^{\dagger}dg
\ket{\xi} = \frac{\bra{\Phi _0}P \ket{\Phi _0}}{\Tr P} P \ket{\xi},
\end{equation}
that is, ${\rm Ran}(P)$ is included by the eigenspace of $\int _G
\pi (g) \ket{\Phi _0}\bra{\Phi _0} \pi (g)^{\dagger} dg$ with an
eigenvalue $\frac{\bra{\Phi _0}P\ket{\Phi _0}}{\Tr P} $. Thus, since
we can see Eq. (\ref{equation 1005}) as a part of spectral
decomposition of a positive operator $\int _G \pi (g) \ket{\Phi
_0}\bra{\Phi _0} \pi (g)^{\dagger}dg $, we can conclude $\mathcal{P}
_{P ^{\perp}}(\ket{\Phi _0}\bra{\Phi _0}) \ge 0$. Therefore, by the
lemma \ref{group representation}, we derive Eq.(\ref{equality of
distance like measure 2}). \hfill $\square$
\end{Proof}


The sufficiency condition of Lemma \ref{group representation 2}
now depends only on $\B_2(\Hi )^{G}$ and a state $\frac{P}{\Tr P}$.
That is, if we know the structure of the representation $(\pi , \Hi )$,
we can check Lemma \ref{group representation 2} without knowing a
closest product state $\ket{\Phi _0}$. Actually, by means of Lemma
\ref{group representation 3}, Lemma \ref{group representation 2} can
be rewritten in a simpler form which is described only in terms of
properties of the group representation $(\pi, \Hi)$ of $G$ as follows.
\begin{Theorem} \label{group representation main}
A projection state $\frac{P}{\Tr P}$ on $\Hi = \otimes _{i=1}^m \Hi
_i$ satisfies,
\begin{equation}
\log _2 (1+R_g(\frac{P}{\Tr P})) = E_R(\frac{P}{\Tr P}) =
G(\frac{P}{\Tr P})-\log_2 \Tr P,
\end{equation}
if there exists a compact topological group $G$ and a finite
dimensional unitary representation $(\pi , \Hi)$ such that $\pi (g
)$ is a local unitary transformation for all $g \in G$, and $(\pi
|_{{\rm Ran }P}, {\rm Ran }P)$ is an irreducible representation of
$G$ whose multiplicity (Definition \ref{multiplicity} in Appendix.
A) is one on $(\pi , \Hi)$.
\end{Theorem}
\begin{Proof}
Suppose the assumption in the statement of the theorem is valid.
Similar to the proof of Lemma \ref{group representation 3}, we can
write $(\pi, \Hi)$ in the form of an irreducible representation as
$(\pi, \Hi)=(\bigoplus _{k=1}^K (I _{\mathcal{A}_k} \otimes \pi _k),
\bigoplus _{k=1}^K (\mathcal{A}_k \otimes \mathcal{B}_k))$, where
$(\pi _k, \mathcal{A}_k \otimes \mathcal{B}_k)$ is irreducible for
all $k$, and $(\pi _k, \mathcal{A}_k \otimes \mathcal{B}_k)$ and
$(\pi _{k'}, \mathcal{A}_{k'} \otimes \mathcal{B}_{k'})$ are not
equivalent for all $k \neq k'$. Without losing generality, we
can assume $\mathcal{B} _1 = {\rm Ran}P$ and $\mathcal{A}_1 =
\mathbb{C}$ by the assumption of the theorem. Then, from Lemma
\ref{group representation 3}, by defining $\{ \ket{\alpha _p^k}
\}_{p=1}^{\dim \mathcal{A}_k}$ as an orthonormal basis of
$\mathcal{A}_k$, we can choose $\{ \frac{\ket{\alpha
_p^k}\bra{\alpha _q^k}\otimes I_{\mathcal{B}_k} }{\sqrt{\dim
\mathcal{B}_k}} \}_{p,q,k}$
 as an orthonormal basis of the Hilbert Schmidt subspace $\B _2(\Hi )^{G}$.
Since $P = \ket{\alpha _1^1}\bra{\alpha _1^1} \otimes I_{\mathcal{B}
_1}$, (note that this $\otimes $ is not the tensor product related
to the entanglement of $P$, that is, $P$ is not a ``separable state''),
$\B ^G _{P^{\perp}} \stackrel{\rm }{=} \{ \rho \in \B _2(\Hi )^G
|\Tr P \rho =0 \}$ can be spanned by $\{ \frac{\ket{\alpha
_p^k}\bra{\alpha _q^k}\otimes I_{\mathcal{B}_k} }{\sqrt{\dim
\mathcal{B}_k}} \}_{p \ge 1, q \ge 1,  k \ge 2}$. Thus, suppose
$\rho \in \Hi \otimes \Hi ^{\dagger}$ is in $\B ^G _{P^{\perp}}$.
Then, $\rho $ can be decomposed only by $\{ \frac{\ket{\alpha
_p^k}\bra{\alpha _q^k}\otimes I_{\mathcal{B}_k} }{ \sqrt{\dim
\mathcal{B}_k}} \}_{p \ge 1, q \ge 1,  k \ge 2}$. Since
$\mathcal{B}_k \perp \mathcal{B}_1$ for all $k \ge 2$, $\rho
\ket{\Psi} = 0 $ for all $\ket{\Psi} \in {\rm Ran} P = \mathcal{B}_1
$. Therefore from Lemma \ref{group representation 2},
Eq.(\ref{equality of distance like measure 2}) holds. \hfill
$\square $
\end{Proof}

In the above proof, we derived Theorem \ref{group representation
main} from Lemma \ref{group representation 2}. However, in
this process we lost no generality; that is, the sufficient conditions of Lemma \ref{group
representation 2} and Theorem \ref{group representation main} are
equivalent. This fact can be seen as follows. Suppose the sufficient
condition of Lemma \ref{group representation 2} is valid. Since $P
\in \B _2(\Hi )^G$, from Lemma \ref{group representation extra 1}
we can see that without losing generality, $P$ can be written down as $P = \sum
_{k=1}^{K_0} P_k \otimes I_{\mathcal{B} _k }$, where $P_k \in \B _2
(\mathcal{A}_k)$ is a non-zero projection, and $K_0 \le K$. First let
us assume $K_0 > 1$. Then, by defining $\sigma \in \B _2 (\Hi )$ and
$\ket{\xi } \in \Hi$ as
\begin{eqnarray*}
\sigma &=& - \left (   \frac{\sum _{k=2}^{K_0}\dim P_k  \dim
\mathcal{B}_k }{\dim P_1  \dim \mathcal{B}_1} \right )P_1 \otimes
I_{\mathcal{B}_1}
+ \sum _{k=2}^{K_0} P_k \otimes I_{\mathcal{B}_k} \\
\ket{\xi} &=& \ket{\alpha } \otimes \ket{\beta},
\end{eqnarray*}
where $\ket{\alpha} \in {\rm Ran}(P_1)$ and $ \ket{\beta} \in
\mathcal{B}_1$, we derive $\Tr P \sigma =0$ and $\sigma \ket{\xi} =
- \frac{\sum _{k=2}^{K_0}\dim P_k  \dim \mathcal{B}_k }{\dim P}
\ket{\xi} \neq 0$; This contradicts the sufficient condition in
Lemma \ref{group representation 2}. Thus, $K_0 = 1$ and ${\rm
Ran}(P) \in \mathcal{A}_1 \otimes \mathcal{B}_1$. Let us now assume
$\dim \mathcal{A}_1 \ge 2$; that is, there exists another equivalent
representation with $(\pi |_{{\rm Ran} P }, {\rm Ran } P )$ in
$(\pi, \Hi)$. In this case, we can write down $P = \ket{\alpha
_1^1}\bra{\alpha _1^1} \otimes I_{\mathcal{B}_1}$ by using $\{
\ket{\alpha _p^k }\}_{p=1}^{\dim \mathcal{A}}$ as an orthonormal
basis of $\mathcal{A}_k$. However, in this case, $\mathcal{A}_1$ is
spanned by $\{ \ket{\alpha _p^1} \}_{p=1}^{d_1}$ for $d_1 \ge 2$. We
define $\sigma \stackrel{\rm def}{=} \sum _{pq} a_{pq} \ket{\alpha
_p^1}\bra{\alpha _q^1} \otimes I_{\mathcal{B}_1}$ with $a_{11} =0 $
and $a_{pq} \neq 0$ ($\forall (p,q) \neq (1,1) $). Then, although
$\sigma \in \B _2(\Hi )^G$, $\sigma \ket{\alpha _1^1} \otimes
\ket{\psi} = \sum _{p \ge 2 } a_{p1}\ket{\alpha _p^1}\bra{\alpha
_1^1} \otimes I_{\mathcal{B}_1} \neq 0$, where $\ket{\psi } \in
\mathcal{B}_1$. This also contradicts the sufficient condition of
Lemma \ref{group representation 2}. Thus, if the sufficient
condition of Lemma \ref{group representation 2} is valid, then,
$(\pi |_{{\rm Ran} P }, {\rm Ran } P )$ is an irreducible
representation of $G$ with multiplicity one on $(\pi, \Hi)$. That
is, the sufficient condition in Lemma \ref{group representation 2}
is equivalent to the sufficient condition in Lemma \ref{group
representation main}.

Finally, we rewrite the above theorem for pure states.
\begin{Theorem} \label{group representation main for pure}
A pure state $\ket{\Psi}$ on $\Hi = \otimes _{i=1}^m \Hi _i$
satisfies,
\begin{equation}
\log _2 (1+R_g(\ket{\Psi })) = E_R(\ket{\Psi }) = E_g(\ket{\Psi}),
\end{equation}
if there exists a compact topological group $G$ and a finite
dimensional unitary representation $(\pi , \Hi)$ such that $\pi (g
)$ is a local unitary transformation for all $g \in G$, and $(\pi
|_{{\rm Ran }\ket{\Psi }\bra{\Psi} }, {\rm Ran
}\ket{\Psi}\bra{\Psi})$ is an irreducible representation of $G$
whose multiplicity is one on $(\pi , \Hi)$.
\end{Theorem}
Thus, if a pure state possesses an enough group symmetry,
the values of all the three distance like measures of
entanglement coincide.
Note that, by means of Theorem \ref{group representation for pure} and Theorem \ref{group representation main for pure},
the results of stabilizer states and symmetric basis states in Section \ref{SEC: Stabilizer and Symm} can be
easily recovered.


\begin{thebibliography}{00}
\bibitem{NoteSupport} The support of a state $\rho$,
with eigen-decomposition $\rho = \sum _i \alpha _i \ket{i}\bra{i}$ is given by $P=\sum _i \ket{i}\bra{i}$.
\bibitem{NoterhoOEW} Note that in the origional definition \cite{T02} $\mathcal{M}$ is the set of
witnesses with trace one. Here we take the slightly extended definition as in \cite{B05} to any
conpact set.
\end{thebibliography}

\end{document}